\documentclass[twocolumn,epjc3]{svjour3}          

\RequirePackage[T1]{fontenc}

\smartqed  

\usepackage{newtxtext}
\usepackage{newtxmath}
\usepackage{wasysym}
\usepackage[english]{babel}
\RequirePackage{graphicx}
\RequirePackage{flushend}
\RequirePackage[numbers,sort&compress]{natbib}
\RequirePackage[colorlinks,citecolor=blue,urlcolor=blue,linkcolor=blue]{hyperref}
\journalname{Eur. Phys. J. C}

\begin{document}
\title{Polarimetry for measuring the vacuum magnetic birefringence with quasi-static fields: a systematics study for the VMB@CERN experiment.}

\author{G.~Zavattini\thanksref{addr1,addr2}
        \and
        F.~Della~Valle\thanksref{e1,addr3,addr4}
        \and
        A.M.~Soflau\thanksref{addr2}
        \and
        L.~Formaggio\thanksref{addr2}
        \and
        G.~Crapulli\thanksref{addr2}
        \and
        G.~Messineo\thanksref{addr1}
        \and
        E.~Mariotti\thanksref{addr3,addr4}
        \and
        \v{S}.~Kunc\thanksref{addr5}
        \and
        A.~Ejlli\thanksref{addr6}
        \and
        G.~Ruoso\thanksref{addr7}
        \and
        C.~Marinelli\thanksref{addr3,addr4}
        \and
        M.~Andreotti\thanksref{addr1}
}

\thankstext{e1}{e-mail: federico.dellavalle@unisi.it}

\institute{INFN, Sez.~di Ferrara, via G.~Saragat 1, Edificio C, 44122 Ferrara, Italy\label{addr1}
        \and
        Dip.~di Fisica e Scienze della Terra, Universit\`a di Ferrara, via G.~Saragat 1, Edificio C, 44122 Ferrara, Italy\label{addr2}
        \and
        Dip.~di Scienze Fisiche, della Terra e dell'Ambiente, Universit\`a di Siena, via Roma 56, 53100 Siena, Italy\label{addr3}
        \and
        INFN, Sez.~di Pisa, largo B.~Pontecorvo 3, 56127 Pisa, Italy\label{addr4}
        \and
        Technical University of Liberec, Studentsk\'a 1402/2, 46117 Liberec, Czech Republic\label{addr5}
        \and
        School of Physics and Astronomy, Cardiff University, Queen's Building, The Parade, Cardiff CF24 3AA, United Kingdom\label{addr6}
        \and
        INFN, Laboratori Nazionali di Legnaro, viale dell'Universit\`a 2, 35020 Legnaro (PD), Italy\label{addr7}
}

\date{Received: \today / Accepted: date}
\titlerunning{The VMB@CERN experiment. A feasibility study.}
\maketitle

\begin{abstract}
We present an experimental systematics study of a polarimetric method for measuring the vacuum magnetic birefringence based on a pair of rotating half-wave plates. The presence of a systematic effect at the same frequency as the sought for magneto-optical effect inhibits the use of strictly constant magnetic fields. We characterise this systematic, discuss its origin and propose a viable workaround.
\end{abstract}

\section{Physics background}

Classical Electrodynamics in vacuum is a linear theory in which the superposition principle holds. In vacuum, in the absence of sources, Maxwell's equations allow wave propagation with velocity $c=1/\sqrt{\epsilon_0\mu_0}$. However, almost a century ago the classical theory of radiation was radically modified by Dirac's relativistic wave equation of the electron \cite{Dirac1928}, which brought along an infinite number of states of negative energy. These states, according to Pauli's principle, are occupied by an infinite number of electrons forming an undetectable background. An electron in a negative energy state could absorb a quantum of light and manifest itself as a real positive-energy electron, leaving behind a ``hole'' in the background; a hole that (before the discovery of the positron) was recognised to play the role of an ``anti-electron'' \cite{Dirac1931}. It was also noted that the energy conservation law could be violated in the short-lived ``intermediate state'' before the annihilation of the particle/anti-particle pair and the reappearance of the quantum of light \cite{Dirac1930}: vacuum fluctuations were hence possible. In this picture it was no longer ``possible to separate processes in the vacuum from those involving matter since electromagnetic fields can create matter if they are strong enough. Even if they are not strong enough to create matter they will, due to the virtual possibility of creating matter, polarise the vacuum and therefore change the Maxwell's equations'' \cite{Heisenberg1936}. ``The light in its passage through the electromagnetic fields will thus behave as if the vacuum took on a dielectric constant that differs from unity as a result of the action of the fields'' \cite{Weisskopf1936}. It was also clear that the new Electrodynamics that was being born had to be no longer linear: photon-photon interaction in vacuum was made possible mediated by electron-positron pairs \cite{Halpern1933}. A lowest order correction to the classical effective Lagrangian density of the electromagnetic field in vacuum taking into account electron-positron pairs was then written for fields slowly varying in space and time \cite{Heisenberg1936,Weisskopf1936,Euler1935,Euler1936}:
\begin{eqnarray}
\nonumber
{\cal L}&=&\frac{1}{2\mu_{\rm 0}}\left(\frac{E^{2}}{c^{2}}-B^{2}\right)+\\
&+&\frac{A_{e}}{\mu_{\rm 0}}\left[\left(\frac{E^{2}}{c^{2}}-B^{2}\right)^{2}+7\left(\frac{\vec{\mathbf{E}}}{c}\cdot\vec{\mathbf{B}}\right)^{2}\right],
\label{Lagrangian}
\end{eqnarray}
where the first term is the classical Lagrangian density of the electromagnetic field and in the correction term the fourth power of the fields appears, allowing light-light interaction. In the correction amplitude
 \[
A_e=\frac{2}{45\mu_{0}}\frac{\alpha^2\mathchar'26\mkern-10mu h^{3}}{m_e^4c^5}=1.32\times 10^{-24} {\rm{~T}}^{-2}
\]
the mass $m_e$ of the electron, the reduced Planck's constant $\mathchar'26\mkern-10mu h$ and the fine structure constant $\alpha$ appear. One should note the smallness of $A_e$ making nonlinear effects hard to be detected in a laboratory measurement. This correction has been confirmed in the framework of Quantum Electrodynamics (QED) \cite{Karplus1950,Schwinger1951}.

Quantum Electrodynamics is one of the most tested theories of physics \cite{Kinoshita1990,Karshenboim2005}, with experiments spanning from atomic physics to high energy phenomena. At high energy the ATLAS and CMS experiments have observed Light-by-Light scattering in Lead-Lead peripheral collisions \cite{ATLAS,CMS}. The same physics has been tackled also with pulsed lasers \cite{Moulin,Sarazin,DiPiazza,Karbstein}. However, one of the predictions of equation~\eqref{Lagrangian}, the vacuum magnetic birefringence (VMB), has never been validated in a laboratory test at low energies, although a hint of this effect from optical astrophysical data has been published \cite{Mignani}. Birefringence is an optical property of anisotropic materials, consisting in the dependence of the index of refraction on the polarisation direction of light with respect to the axes of the system. It is quite common in crystalline solids, whereas in isotropic materials it can be induced by stress or by electric or magnetic fields, with the field direction defining the optical axis of the system.

In vacuum, from the Lagrangian of equation~\eqref{Lagrangian}, constitutive equations can be derived for $\vec{\mathbf{D}}$ and $\vec{\mathbf{H}}$ \cite{Weisskopf1936,Euler1935,Euler1936}:
\begin{eqnarray*}
\vec{\mathbf{D}}&=&\frac{\partial{\cal L}}{\partial\vec{\mathbf{E}}}=\\
&=&\displaystyle\epsilon_0\left[\vec{\mathbf{E}}+4A_e\left(\frac{E^2}{c^2}-B^2\right)\vec{\mathbf{E}}+14A_e\left(\vec{\mathbf{E}}\cdot\vec{\mathbf{B}}\right)\vec{\mathbf{B}}\right],
\end{eqnarray*}
\begin{eqnarray*}
\mu_0\vec{\mathbf{H}}&=&-\mu_0\frac{\partial{\cal L}}{\partial\vec{\mathbf{B}}}=\\
&=&\displaystyle\vec{\mathbf{B}}+4A_e\left(\frac{E^2}{c^2}-B^2\right)\vec{\mathbf{B}}-14A_e\left(\frac{\vec{\mathbf{E}}}{c}\cdot\vec{\mathbf{B}}\right)\frac{\vec{\mathbf{E}}}{c}.
\end{eqnarray*}
These equations describe a nonlinear anisotropic medium. We are interested in the case in which a light wave travels through an external magnetic field $\vec{\mathbf{B}}_{\rm ext}$. Then the electric field is the electric field of the light $\vec{\mathbf{E}}=\vec{\mathbf{E}}_\gamma$ whereas the magnetic field is the sum $\vec{\mathbf{B}}=\vec{\mathbf{B}}_\gamma+\vec{\mathbf{B}}_{\rm ext}$. We also suppose that $B_{\rm ext}\gg B_\gamma=E_\gamma/c$ and that the propagation direction of the wave is perpendicular to the external field. We are interested in the dynamic fields; to first order in the fields of the light the above equations become
\begin{eqnarray*}
&&\vec{\mathbf{D}}_\gamma=\displaystyle\epsilon_0\left[\vec{\mathbf{E}}_\gamma-4A_eB^2_{\rm ext}\vec{\mathbf{E}}_\gamma+14A_e\left(\vec{\mathbf{E}}_\gamma\cdot\vec{\mathbf{B}}_{\rm ext}\right)\vec{\mathbf{B}}_{\rm ext}\right],\\
&&\mu_0\vec{\mathbf{H}}_\gamma=\displaystyle\vec{\mathbf{B}}_\gamma-4A_eB_{\rm ext}^2\vec{\mathbf{B}}_\gamma-8A_e\left(\vec{\mathbf{B}}_\gamma\cdot\vec{\mathbf{B}}_{\rm ext}\right)\vec{\mathbf{B}}_{\rm ext}.
\end{eqnarray*}
We distinguish two cases: one in which $\vec{\mathbf{E}}_\gamma\parallel\vec{\mathbf{B}}_{\rm ext}$ and the second in which $\vec{\mathbf{E}}_\gamma\perp\vec{\mathbf{B}}_{\rm ext}$. The vacuum is then found to be a birefringent uniaxial medium whose electric and magnetic susceptibilities and indices of refraction are \cite{Toll1952,Erber1961,Klein1964,Baier1967,Bialynicka1970,Adler1971,GrassiStrini1975}
\begin{eqnarray*}
&\epsilon_\parallel=1+10A_eB_{\rm ext}^2,\qquad&\epsilon_\perp=1-4A_eB_{\rm ext}^2,\\
&\mu_\parallel=1+4A_eB_{\rm ext}^2,\qquad&\mu_\perp=1+12A_eB_{\rm ext}^2,\\
&n_\parallel=1+7A_eB_{\rm ext}^2,\qquad&n_\perp=1+4A_eB_{\rm ext}^2,
\end{eqnarray*}
and
\begin{equation}
\Delta n_{\rm QED}=n_\parallel-n_\perp=3A_eB_{\rm ext}^2\simeq 4\times10^{-24}B_{\rm ext}^2.
\label{VMB}
\end{equation}
This equation has the same structure as a magnetically induced birefringence in condensed matter and gases \cite{Rizzo1997}, suggesting again a picture of the quantum vacuum as a peculiar material medium. Vacuum magnetic birefringence in equation~\eqref{VMB} is the dominant effect on the propagation of light in a vacuum. A magnetic birefringence is also associated to hitherto hypothetical axion-like and milli-charged particles \cite{Maiani1986,Gies2006}. In general both could also generate a magnetic dichroism (dependence of the imaginary part of the complex index of refraction on the polarisation direction of light), detectable as a rotation of the polarisation direction. 

\begin{figure}[bht]
\begin{center}
\includegraphics[width=8.4cm]{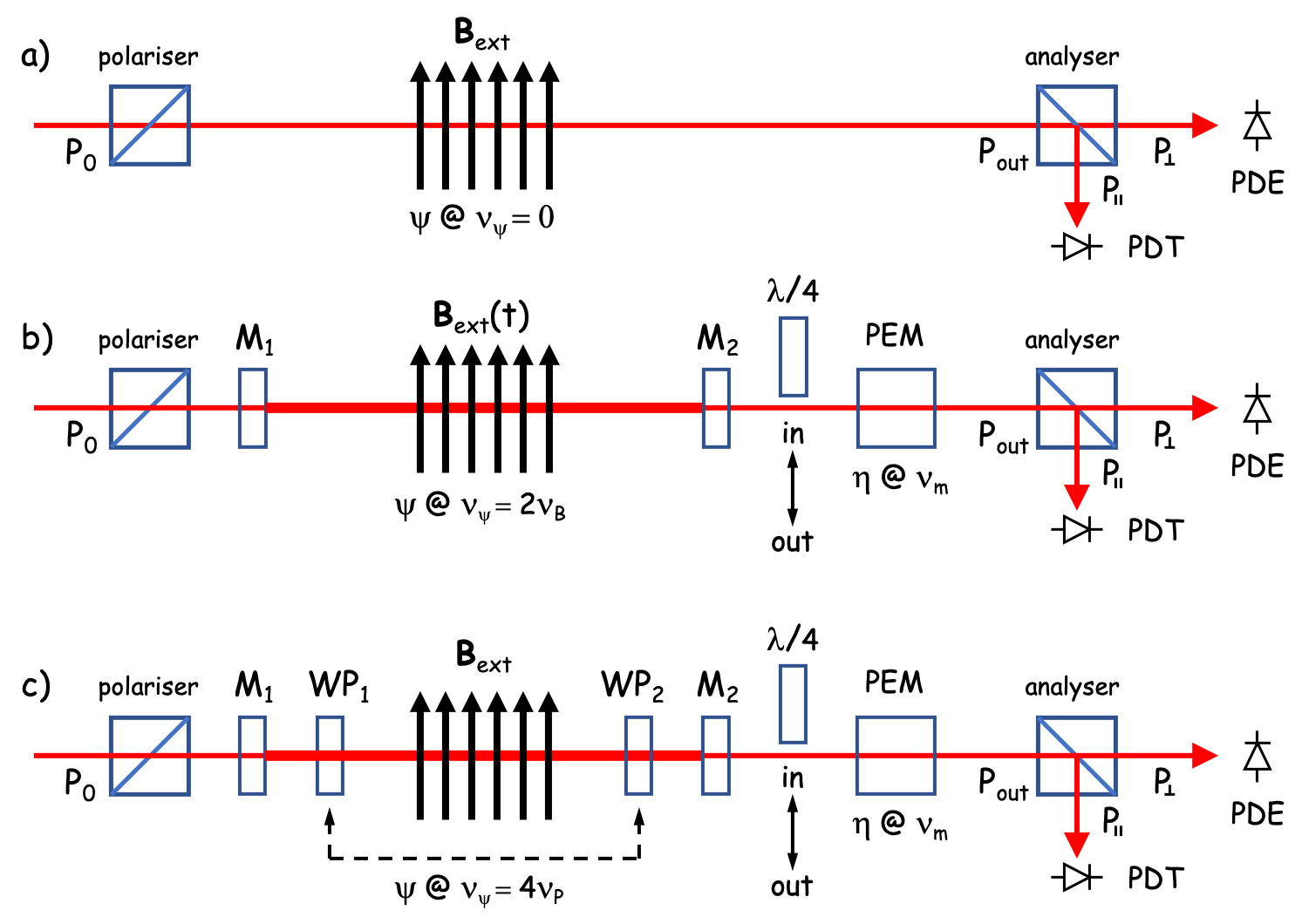}
\end{center}
\caption{Polarimetric schemes: PDE: Extinction Photodiode; PDT: Transmission Photodiode; M$_{1,2}$: cavity mirrors; $\lambda/4$: quarter-wave plate; PEM: Photo-Elastic Modulator; WP$_{1,2}$: rotating half-wave plates. The quarter-wave plate is inserted when rotation measurements are performed.}
\label{Polarimetry}
\end{figure}

Traversing a birefringent medium, the components of the electric field of light along the birefringence axes acquire a phase difference $\Delta\varphi$. A linearly polarised light beam acquires an ellipticity\protect\footnote{Ratio (with sign) of the minor to the major axis of the ellipse described by the electric vector of the light. The sign distinguishes between the two rotation directions of the electric field around the ellipse.}. It can be shown that, for small phase differences, the ellipticity is
\begin{equation}
\psi(\phi)=\sin2\phi\int\frac{\pi\,\Delta n}{\lambda}\,dz=\psi_0\sin2\phi=\frac{\Delta\varphi}{2}\sin2\phi
\label{Ellipticity}
\end{equation}
where the integral is performed along the light path and $\phi$ is the angle between the magnetic field direction and the initial polarisation. To measure an acquired ellipticity one has then to measure the electric field of the light in the direction orthogonal to the initial polarisation. In the case of the vacuum, the ellipticity values attainable in a laboratory measurement are quite small. For $\lambda=1064$~nm, $B_{\rm ext}=2.5$~T and a light path in the magnetic field $L_B=1.64$~m (parameters of the PVLAS experiment \cite{PVLAS2016,PVLAS2020,Universe})
\[
\psi_0^{\rm (PVLAS)}=\frac{\pi\,\Delta nL_B}{\lambda}=1.2\times10^{-16}.
\]
An analyser set to maximum extinction (see figure~\ref{Polarimetry}a) would transmit the power $P_\perp=P_{\rm out}\,\psi_0^2\sin^22\phi\approx P_\parallel\,\psi_0^2\sin^22\phi$ where now the subscripts $\perp$ and $\parallel$ refer to the initial polarisation direction. In the case of vacuum magnetic birefringence this intensity is far less than the extinguished power $P_{\rm out}\sigma^2$ of any existing pair of polarisers, imposing a more sophisticated approach.

\section{Polarimetric method}

\subsection{Modulation polarimetry}

The modern history of the measurement of the vacuum magnetic birefringence in the optical range began in 1979 with the seminal paper by E. Iacopini and E. Zavattini \cite{Iacopini1979}. The proposed method linearises the signal by modulating the effect and summing to it a known modulated ellipticity, as seen in figure~\ref{Polarimetry}b. In this scheme the signal is found in a Fourier analysis of the extinguished power at the frequencies sum and difference of the two modulation frequencies (heterodyne detection). If one of the two ellipticities is static, one has a homodyne detection. Following the guidelines of the 1979 paper, several experimental efforts have been set up in the attempt to measure the tiny vacuum magnetic birefringence \cite{Iacopini1981,BFRT1993,PVLAS1998,PVLAS2008a,PVLAS2008b,Q&A2004,Q&A2010,BMV2004,BMV2014,OSQAR2006,PVLAS2013,PVLAS2016,PVLAS2020,OVAL2017,Universe}, without success. The modulation of the effect can be obtained either by varying the magnetic field intensity or its direction. To realise this second type of modulation, the PVLAS experiment \cite{PVLAS1998,PVLAS2008a,PVLAS2008b,PVLAS2013,PVLAS2016,PVLAS2020,Universe} rotated continuously the magnets at a frequency $\nu_B$. In this case, according to equation~\eqref{Ellipticity}, the effect was modulated at a frequency $\nu_\psi=2\nu_B$. Neglecting the polariser extinction $\sigma^2$, the power in the extinguished beam is now
\begin{equation}
P_{\perp}=P_{\rm out}\left[\psi(t)+\eta(t)\right]^2\approx P_\parallel\left[\eta^2(t)+2\eta(t)\psi(t)\right],
\label{ExtinguishedBeam}
\end{equation}
where $\eta(t)$ is a known modulated ellipticity. In the PVLAS experiment this modulation was obtained by employing a Photo-Elastic Modulator (PEM) (see figure~\ref{Polarimetry}b). In the following we will assume that $\eta(t)=\eta_0\cos(2\pi\nu_{\rm m}t+\phi_{\rm m})$, and $\psi(t)=\psi_0\cos(2\pi\nu_\psi t+\phi_\psi)$. Therefore in the Fourier spectrum of the extinguished power the sought for ellipticity can be extracted from the amplitude of the sidebands of the modulation frequency $\nu_{\rm m}$ at $\nu_{\rm m}\pm\nu_\psi$. Using two lock-in amplifiers to demodulate the extinguished power at the frequencies $\nu_{\rm m}$ and $2\nu_{\rm m}$, the measured ellipticity $\psi_0$ is given by \cite{PVLAS2016,PVLAS2020,Universe}
\begin{equation}
\psi_0=\psi(\nu=\nu_\psi)=\frac{P_{\nu_{\rm m}}(\nu_\psi)}{P_{2\nu_{\rm m}}({\rm DC})}\frac{\eta_0}{4}.
\label{Psi0}
\end{equation}
The same equation is used in a rotation measurement (quarter-wave plate inserted). We note explicitly that the above equation holds for any frequency $\nu$ of the demodulated spectrum. In the following it will be used to study the noise of the polarimeter in a wide frequency band.

A common point to all the experiments attempted so far has been the presence of a light path amplifier, realised in all but one case by employing a Fabry-Perot cavity whose finesse the experiments strived to increase. Indeed the PVLAS experiment \cite{PVLAS2016,PVLAS2020,Universe} set the best limits on the VMB employing a cavity with a finesse $F_{\rm PVLAS}\approx7\times10^5$ and a path-amplification factor $N_{\rm PVLAS}=2F/\pi\approx4.5\times10^5$ coupled with two permanent dipole magnets that were rotated to modulate the effect. The experiment ended in 2018 after reaching a noise floor less than a factor ten from the predicted QED value, limited by the existence of an intrinsic $\sim1/f$ noise source due to the high reflectivity mirrors of the Fabry-Perot cavity \cite{PVLAS2020,IntrinsicNoise}. The measurements showed that, for high finesse cavities, the signal-to-noise ratio did not improve with the finesse, frustrating the rush for higher and higher finesses and limiting the sensitivity.

In a polarimetric measurement the limit ellipticity noise is the shot noise, whose peak spectral density is
\[
S^{(\rm shot)}_\psi=\sqrt{\frac{e}{qP_{\rm out}}},
\]
where $e$ is the elementary charge and $q$ is the quantum efficiency of the extinction photodiode. This number determines the minimum time $T_{\rm min}$ required to reach a unitary signal-to-noise ratio in the measurement of an ellipticity $\psi_0$:
\begin{equation}
T_{\rm min}=\left[\frac{S^{(\rm shot)}_\psi}{\psi_0}\right]^2.
\label{Tmin}
\end{equation}
In the case of the PVLAS experiment, with $S^{(\rm shot)}_\psi\approx5\times10^{-9}/\sqrt{\rm Hz}$, this minimum time was \cite{PVLAS2016,PVLAS2020,Universe}
\[
T_{\rm min}=\left[\frac{S^{(\rm shot)}_\psi}{N_{\rm PVLAS}\,\psi_0^{\rm (PVLAS)}}\right]^2\approx10^4~{\rm s}.
\]
Unfortunately, the observed PVLAS intrinsic mirror noise was almost two orders of magnitude larger than the shot noise at the attainable modulation frequency $\nu_\psi\approx20$~Hz \cite{PVLAS2016,PVLAS2020,Universe}. With respect to the shot-noise-limited case the measurement time was a factor $\sim10^4$ larger, making impossible the continuation of the experiment.

The only way to get around this issue is to increase the magnetic field intensity and length, the relevant parameter being the product $B_{\rm ext}^2L_B$. By employing a LHC main bending dipole \cite{Rossi2003} this parameter could reach almost 1200~T$^2$m, a factor 100 larger than in the PVLAS experiment. For light with $\lambda=1064$~nm, such a magnet would generate an ellipticity
\begin{equation}
\psi_0^{\rm (CERN)}=\frac{3\pi A_eB_{\rm ext}^2L_B}{\lambda}=1.4\times10^{-14}.
\label{ell_cern}
\end{equation}
Unfortunately large superconducting magnets cannot be rotated and can be modulated only at frequencies $\lesssim10$~mHz (see §\ref{Induttanza} below). In order to keep the original 1979 measurement scheme, a new polarisation modulation method has been proposed \cite{RotatingWavePlates} which could be the heart of a new experiment \cite{CERN_LOI,PBC2020}. The method allows the use of (quasi) static magnetic fields with the ellipticity signal modulated by a polarisation rotation resulting in $\nu_\psi\gtrsim10$~Hz. The next section presents the method which is later discussed in the light of the proof of principle tests that have been carried out. We anticipate here that, as a result of the present experimental study, a low-frequency modulation of the magnetic field will be needed.

\subsection{Rotating polarisation}

Given the practical difficulty of modulating either the current or the direction of a large superconducting magnet, the proposed polarimetric scheme envisages to make the polarisation rotate inside the magnet \cite{RotatingWavePlates}. This can be obtained by employing a pair of half-wave plates co-rotating at frequency $\nu_{\rm P}$ (see figure~\ref{Polarimetry}c). The use of a rotating half-wave plate had been already proposed in the framework of the OSQAR experiment \cite{OSQAR2006}. However, in that proposal the wave plate was outside the Fabry-Perot cavity and the polarisation would have been rotating also on the reflecting surface of the dielectric mirrors of the cavity. It is well known that these surfaces are birefringent with phase differences of order $10^{-6}$~rad/reflection \cite{PVLAS2016,PVLAS2020,Bouchiat1982,Carusotto1989,Micossi1993,Brandi1997,Zavattini2006}. This birefringence, modulated at the same frequency as the vacuum magnetic one but many orders of magnitude larger, would show up in the extinguished beam of equation~\eqref{ExtinguishedBeam} as by far the dominant signal. In the newly proposed scheme, on the contrary, the polarisation does not rotate on the mirrors. The first wave plate makes the polarisation rotate at a frequency $2\nu_{\rm P}$ inside the magnetic region, whereas the second one stops the rotation. The effect of the birefringence due to the magnetic field will therefore be modulated at a frequency $\nu_\psi=4\nu_{\rm P}$. The two wave plates need not have their axes aligned: in this scheme both mirrors can easily have their axes aligned to the local polarisation direction, making the effect induced by the mirror birefringence static and minimised. As we have not yet tested the method with a Fabry-Perot cavity, we will not further discuss the birefringence of the cavity mirrors. This subject has been thoroughly investigated with the PVLAS apparatus \cite{PVLAS2016,PVLAS2020,Universe}.
We now want to study the spurious signals and the sensitivity of this new method. As a first point, the phase retardation of the half-wave plates WP$_1$ and WP$_2$ is expected to slightly deviate from $\pi$ by amounts $\alpha_1,\alpha_2\sim$~mrad. It can be shown then that to first order in $\alpha_1$, $\alpha_2$ and $\psi_0$ the light power detected by the photodiode PDE is given by \cite{RotatingWavePlates}
\begin{eqnarray}
\label{ExtinguishedIntensity}
&&P_\perp\approx P_{\rm out}\left\{\eta^2+\,N\eta\Big[\right.2\psi_0\sin(4\phi_{\rm P}+4\phi_1)+\\
\nonumber&&+\left.\alpha_1\sin(2\phi_{\rm P}+2\phi_1)+\alpha_2\sin(2\phi_{\rm P}+2\phi_1-2\Delta\phi)\Big]\right\},
\end{eqnarray}
where $\eta=\eta(t)$ is the known modulated ellipticity, $\phi_{\rm P}=2\pi\nu_{\rm P}t$ is the phase of the rotation, $\Delta\phi=\phi_2-\phi_1$ and $\phi_1$ and $\phi_2$ are arbitrary phases of the angular azimuthal position of the slow axis of the two plates. From this formula one can see that the signal of the magnetic birefringence of vacuum appears at a frequency $\nu_{\rm m}\pm4\nu_{\rm P}$, whereas the signals due to imperfect wave plates should come at $\nu_{\rm m}\pm2\nu_{\rm P}$. However, as will be seen below, the $\alpha's$ are not to be regarded as constant quantities, as they depend on alignment and on optical and geometrical properties of the wave plates, generating further time dependencies in equation~\eqref{ExtinguishedIntensity}.

Before tackling the argument of the systematics, we want to show that the method promises to give a signal-to-noise ratio such that the measurement of the vacuum magnetic birefringence can be carried out in a reasonable time. On the signal side, we begin noticing that the presence of intra-cavity optical elements limits the attainable finesse. In the case we are discussing, this limit is given by the antireflective coating of the wave plates. The reflectivity of commercial AR-coatings is $R_{\rm AR}\lesssim0.1\%$, but one may hope to obtain at least a factor two better \cite{FiveNine}. Identifying the cavity losses with this residual reflectivity, taking into account the four surfaces of the wave plates and neglecting the transmittance of the mirrors, a conservative estimate for the amplification factor $N_{\rm HWP}$ is
\[
N_{\rm HWP}=\frac{2F}{\pi}\approx\frac{2}{4R_{\rm AR}}=500.
\]
The experience of the PVLAS experiment reveals that the shot noise has been reached in low finesse cavities \cite{PVLAS2016,PVLAS2020,Universe}. Assuming that this will be true also in the present situation, one can calculate the minimum laser power $P_{\rm out,min}$ needed to maintain the integration time of equation~\eqref{Tmin} down to $T_{\rm min}\approx10^5$~s (about one day):
\[
P_{\rm out,min}=\frac{e}{q\left[N_{\rm HWP}\,\psi_0^{\rm (CERN)}\right]^2T_{\rm min}}\approx50~{\rm mW},
\]
a perfectly manageable value.

We note explicitly that equation~\eqref{ExtinguishedIntensity} is written with the assumption that $N\alpha_{1,2}\ll1$, so as to have a well defined linear polarisation between the cavity mirrors. In our case $N\approx500$, hence $\alpha_{1,2}\lesssim10^{-4}$. As this cannot be guaranteed by current commercial wave plates (as will be verified in the following), and given the temperature dependence of $\alpha$ \cite{TemperatureDependence}, we are developing a closed-loop temperature control system to keep $\alpha$ within our requirements. 

\subsection{Systematics}
\label{Systematics}

\subsubsection{The effect of angular alignment}
\label{AlignmentSystematics}

\begin{figure}[bht]
\begin{center}
\includegraphics[width=6cm]{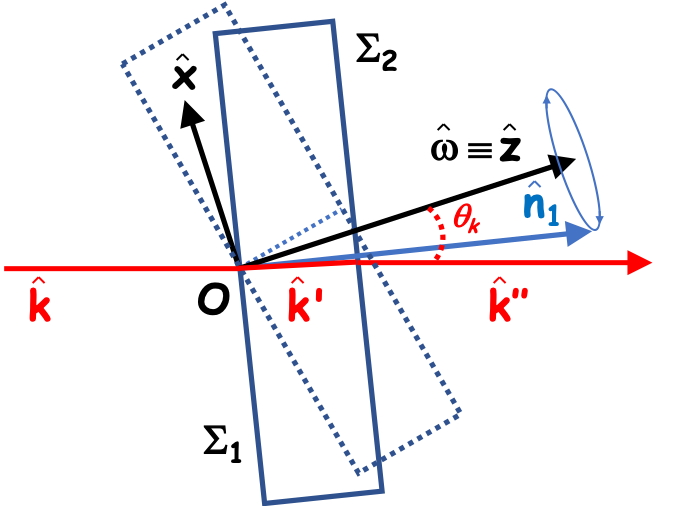}
\end{center}
\caption{Geometry of a rotating wave plate traversed by a light beam. The appropriate phase of the rotation has been chosen to have $\hat{\mathbf{n}}_1$ in the same plane as $\hat{\mathbf{k}}$ and $\hat{\mbox{\boldmath$\omega$}}$; the dotted wave plate is drawn at $\Delta\phi=\pi$ with respect to the continuous line scheme.}
\label{Axes}
\end{figure}

Let us begin describing the rotation of a nominally flat wave plate traversed by a laser beam. We consider a fixed rotation axis not orthogonal to the plate and not coinciding with the light beam. Referring to figure~\ref{Axes}, we need to specify three axes:
\begin{enumerate}
\item
$\hat{\mbox{\boldmath$\omega$}}$ is the rotation axis of the optical element; we assume it time independent and defining the $z$ axis, with origin O on the first surface of the wave plate;
\item
$\hat{\mathbf{n}}_1$ and $\hat{\mathbf{n}}_2$ are the local normals to the optical surfaces $\Sigma_1$ and $\Sigma_2$ of the wave plate at the positions where the light beam intersects the two surfaces; 
\item
$\hat{\mathbf{k}}$ is along the incoming light beam forming a small constant angle $\theta_k$ with $\hat{\mbox{\boldmath$\omega$}}$.
\end{enumerate}

We assume for now that $\hat{\mathbf{k}}$ intersects $\hat{\mbox{\boldmath$\omega$}}$ on the first surface of the wave plate in the (still) point O assumed as the origin of the reference system. In this way one is only concerned with the effect of angular misalignments. With the assumptions made, $\hat{\mathbf{n}}_1$ precedes around the $z$ axis forming with it the constant angle $\theta_{n1}$. Due to the rotation, the refracted and exit directions $\hat{\mathbf{k}}'$ and $\hat{\mathbf{k}}''$ oscillate; $\hat{\mathbf{k}}''$ precedes around $\hat{\mathbf{k}}$ describing a truncated cone of semi-aperture approximately equal to the wedge angle $\beta$ of the plate; as the plates can be manufactured with $\beta\approx1~\mu$rad, in the present context this change of direction is negligible and hence $\hat{\mathbf{k}}\parallel\hat{\mathbf{k}}''$ and $\hat{\mathbf{n}}_1\parallel\hat{\mathbf{n}}_2\equiv\hat{\mathbf{n}}$.

The components of the vectors above are:
\begin{eqnarray*}
\hat{\mbox{\boldmath$\omega$}}&=&(0,0,1),\\
\hat{\mathbf{k}}&=&(\sin\theta_k,0,\cos\theta_k),\\
\hat{\mathbf{n}}&=&(\sin\theta_{n}\cos\phi_{\rm P},\sin\theta_{n}\sin\phi_{\rm P},\cos\theta_{n}),
\end{eqnarray*}
where $\theta_n=\theta_{n1}$ and $\hat{\mathbf{n}}$ depends on time through $\phi_{\rm P}(t)$. The incidence angle $\theta_i$ is
\[
\cos\theta_i=\hat{\mathbf{k}}\cdot\hat{\mathbf{n}}=\sin\theta_k\sin\theta_{n}\cos\phi_{\rm P}+\cos\theta_k\cos\theta_{n}.
\]
The light path $L$ inside the plate is written in terms of the refracted angle $\theta_r$:
\begin{equation}
L=\frac{D}{\cos\theta_r}=\frac{D}{\sqrt{1-\displaystyle\frac{\sin^2\theta_i}{n^2}}}\approx D\left(1+\frac{\sin^2\theta_i}{2n^2}\right)
\label{LightPath}
\end{equation}
where $D$ is the thickness of the wave plate. Note that for a non perfect alignment the incidence angle $\theta_i$ depends on $\phi_{\rm P}$ and therefore equation \eqref{LightPath} introduces further azimuthal dependences in the extinguished intensity of equation~\eqref{ExtinguishedIntensity}. In fact, the above expression has, besides a DC term, two Fourier components at $\nu_{\rm P}$ and $2\nu_{\rm P}$ of amplitude
\begin{eqnarray*}
L_{\nu_{\rm P}}&=&\frac{D}{4n^2}\sin2\theta_{n}\sin2\theta_k,\\
L_{2\nu_{\rm P}}&=&\frac{D}{4n^2}\sin^2\theta_{n}\sin^2\theta_k.
\end{eqnarray*}
The small phase retardation errors of such a wave plate will generate spurious signals in equation~\eqref{ExtinguishedIntensity} given by
\begin{equation}
\alpha=\frac{2\pi}{\lambda}\int\Delta n\,dz\pm p\pi=\frac{2\pi\Delta nL}{\lambda}\pm p\pi
\label{Alpha}
\end{equation}
where the integral is performed along the light path inside the wave plate, $\Delta n_{\rm quartz}\approx0.00874$ at $\lambda=1064$~nm \cite{Ghosh1999} and $p$ is an appropriate odd integer. In our set-up the phase errors \eqref{Alpha}, now containing frequency components at $\nu_{\rm P}$ and $2\nu_{\rm P}$, will beat with the $2\phi_{\rm P}$ dependence of equation~\eqref{ExtinguishedIntensity} to give components at $\nu_{\rm P}$, $3\nu_{\rm P}$ and $4\nu_{\rm P}$. This last component is indistinguishable from $\psi_0$ and one must ensure that $\alpha_{2\nu_{\rm P}}/4\ll\psi_0$ or
\[
\frac{\pi\,\Delta n\,D}{8n^2\lambda}\sin^2\theta_{n}\sin^2\theta_k\ll\psi_0^{\rm (CERN)}\approx10^{-14},
\]
where $n\approx1.54$ and we assume $D\sim1$~mm. The condition above is then satisfied if $\theta_k,\theta_{n}\ll3\times10^{-4}~\mbox{\rm rad}\approx1$~arcmin, which seems an easily reachable condition. With $\theta_{n},\theta_k\approx3\times10^{-4}$~rad, the ellipticity signals at $\nu_{\rm P}$ and $3\nu_{\rm P}$ are instead much larger, $\approx5\times10^{-7}$.

\subsubsection{Intrinsic systematics}
\label{IntrinsicSystematics}

With the geometry described above one can study other spurious effects. Let us note first that if the laser beam impinges in O but the alignment is not perfect the beam describes on $\Sigma_2$ a minimal circumference $\Gamma_0$ with radius $r_0$
\[
r_0\approx D(\sin\theta_i-\sin\theta_r)\approx\frac{D}{2}\sin\theta_i\frac{n-1}{n}
\]
where $\theta_i$ varies during rotation between $\theta_k-\theta_{n}$ and $\theta_k+\theta_{n}$. With $\theta_{n},\theta_k\approx3\times10^{-4}$~rad, $r_0\approx5\times10^{-8}$~m, an unmeasurably small value. Hence, if the beam does not impinge on the first surface exactly in O, the impact point of the beam will describe a circle on the first as well as on the second surface of the wave plate. Let us address the points of the two curves with a transverse position vector $\vec{\mathbf{r}}_{\phi_{\rm P}}$ rotating at frequency $\nu_{\rm P}$. As both surfaces $\Sigma_1$ and $\Sigma_2$ suffer from slope errors, the intrinsic thickness $D$ in equation~\eqref{LightPath} becomes a function of the rotation angle $\phi_{\rm P}$: $D=D(\vec{\mathbf{r}}_{\phi_{\rm P}})$ and will contain harmonics of $\phi_{\rm P}$. Furthermore, also $\Delta n$ will depend on $\vec{\mathbf{r}}_{\phi_{\rm P}}$. The transverse gradient of $\Delta nD$ corresponds to the first $\phi_{\rm P}$ harmonic component given by
\begin{eqnarray}
\label{GradientDeltanD}
\left(\Delta n\,D\right)_{\nu_{\rm P}}&=&\vec{\mathbf{r}}_{\phi_{\rm P}}\cdot\vec{\mbox{\boldmath$\nabla$}}(\Delta n\,D)=
\\&=&\vec{\mathbf{r}}_{\phi_{\rm P}}\cdot\left[D\vec{\mbox{\boldmath$\nabla$}}(\Delta n)+\vec{\mbox{\boldmath$\beta$}}\Delta n\right],
\nonumber
\end{eqnarray}
where $\vec{\mbox{\boldmath$\beta$}}=\vec{\mbox{\boldmath$\nabla$}}D$ has amplitude the wedge angle of the plate. A realistic estimate for the positioning error of the laser beam with respect to the point O might be $r_{\phi_{\rm P}}\approx0.1$~mm. Inserting this value in the second term of the above equation~\eqref{GradientDeltanD} with $D\approx1$~mm, $\Delta n=\Delta n_{\rm quartz}=0.00874$ and $\beta\approx10^{-5}$ one finds a contribution to the variation of the optical path difference of $5\times10^{-11}$~m, corresponding to an $\alpha_{\nu_{\rm P}}\approx5\times10^{-5}$~rad. This first harmonic phase error will beat with the $2\nu_{\rm P}$ in equation~\eqref{ExtinguishedIntensity} generating harmonics in the ellipticity at $\nu_{\rm P}$ and $3\nu_{\rm P}$.

Higher order contributions can be envisaged according to the (optical) symmetry of the surface encircled by the beam path of radius $r_{\phi_{\rm P}}$. The further the light beam moves away from O, the larger is $r_{\phi_{\rm P}}$, and hence the larger becomes the first order contribution due to the global character of the term $\vec{\mbox{\boldmath$\beta$}}\Delta n$; this is not necessarily the case for the higher order terms.

In summary, the most general expression for the quantity $\alpha$ of equation~\eqref{Alpha}, which includes both what we referred to as alignment and intrinsic effects, is
\begin{equation}
\alpha(\phi_{\rm P})=\alpha_0+\alpha_{\nu_{\rm P}}\cos\phi_{\rm P1}+\alpha_{2\nu_{\rm P}}\cos2\phi_{\rm P2}+\ldots
\label{DefectSymmetry}
\end{equation}
where, in order to allow for a different phase for each $s$-component,
\begin{equation}
\phi_{{\rm P}s}=\phi_{\rm P}(t)+\phi_{s,0}.
\label{Phi_s,0}
\end{equation}
In the expression \eqref{DefectSymmetry} each term generates a spurious ellipticity component corresponding to a well defined azimuthal symmetry, with constant amplitude coefficients $\alpha_{s\nu_P}$ for fixed $\theta_n$, $\theta_k$ and $r_{\phi_{\rm P}}$. However, below we will discuss a case in which $\alpha_{\nu_P}$ is modulated at the frequency $\nu_P$, thus mimicking an extra second order component in $\alpha(\phi_{\rm P})$. This is of particular interest because a second order dependence of $\alpha(\phi_{\rm P})$ produces a fourth harmonic when inserted in equation~\eqref{ExtinguishedIntensity}. This component proves to be much larger than the one due to a second order intrinsic defect of the wave plate.

\section{Experimental set-up}

\begin{figure}[bht]
\begin{center}
\includegraphics[width=8.4cm]{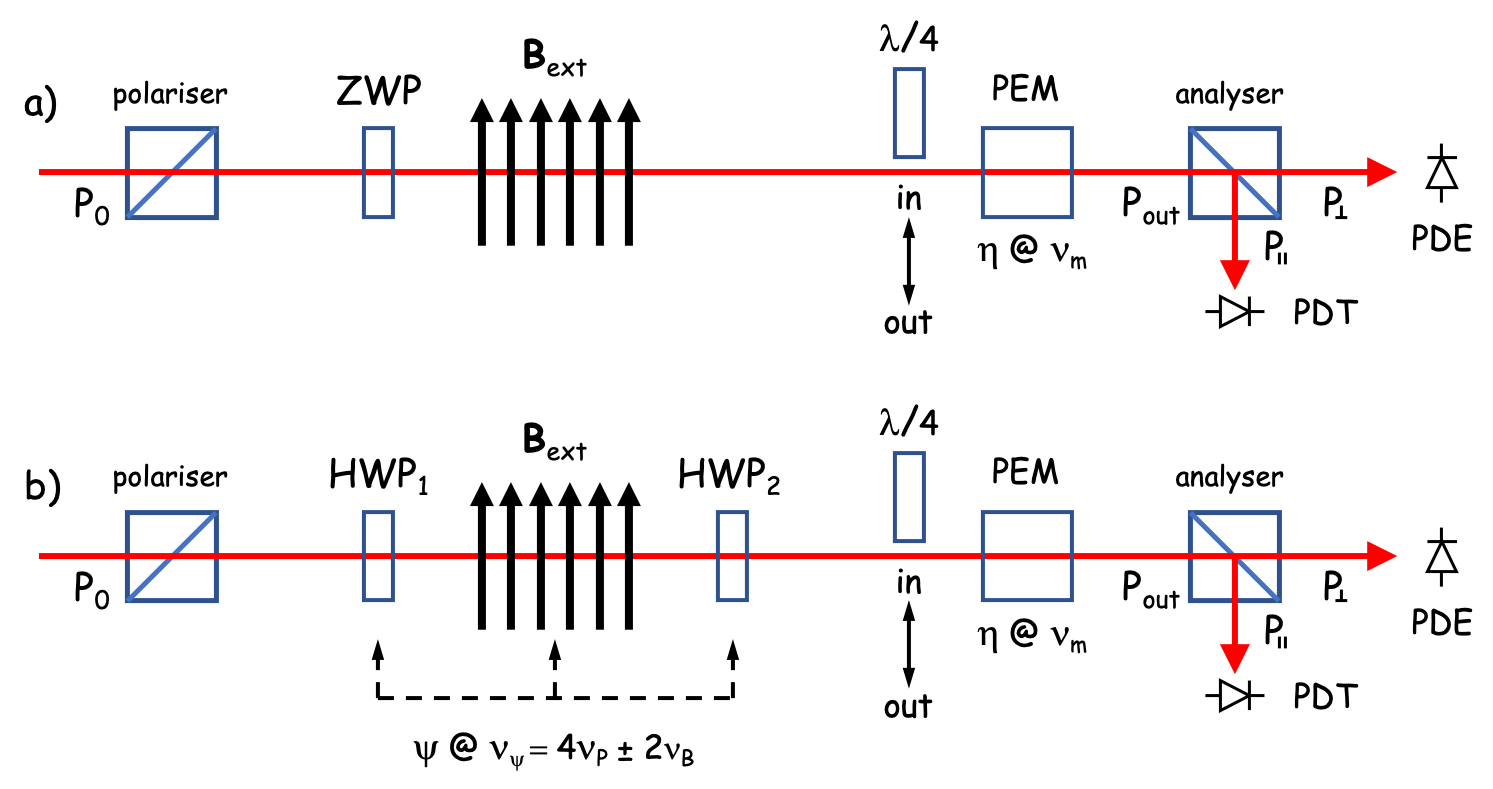}
\end{center}
\caption{Experimental set-ups: PDE: Extinction Photodiode; PDT: Transmission Photodiode; $\lambda/4$: quarter-wave plate; PEM: Photo-Elastic Modulator; ZWP: rotating zero-wave plate; HWP$_{1,2}$: rotating half-wave plates. The quarter-wave plate is inserted when rotation measurements are performed.}
\label{SetUp}
\end{figure}

The two experimental set-ups employed in the present work are shown in figure~\ref{SetUp}. Both are simplified versions (deprived of the cavity mirrors) of the original idea of a polarimeter with two co-rotating half-wave plates \cite{RotatingWavePlates} depicted in figure~\ref{Polarimetry}c. The completion of the polarimetric scheme with a Fabry-Perot cavity will be the object of a future work. 
In the measurements we will present, the light path will be either in air 
or in a controlled atmosphere of a pure gas. The set-up makes use of the two identical rotatable permanent magnet dipoles of the PVLAS experiment \cite{PVLAS2016,PVLAS2020,Universe}. Each of them generates a magnetic field $B_{\rm ext}=2.5$~T over a length $L_B=0.82$~m, with $B_{\rm ext}^2L_B=5.125\pm0.040$~T$^2$m.

Panel a) of figure~\ref{SetUp} features a single nominally neutral rotating optical element (zero-wave plate). Two different zero-wave plates have been used in turn: an uncoated glass plate 1~cm thick (GP) and an AR-coated optical assembly of two crossed half-wave plates (CWP), 1.6~mm thick, with $6.5\,\lambda$ optical path difference per plate. Using these optical elements no modulation of the magneto-optic effect takes place, the set-up being apt to study the mere effects on the polarisation due to the wave plate rotation and the relative systematics, without the complications of the two co-rotating plates and of the rotating polarisation. In this case, in fact, a static magnetic field induces only a constant ellipticity.

In the b) panel of the same figure two co-rotating half-wave plates make the polarisation rotate inside the magnetic region at a frequency $2\nu_{\rm P}$. This modulates the ellipticity induced by the magnetic birefringence of the gas inside the magnets at a frequency $\nu_\psi=4\nu_{\rm P}$. In this set-up the magnets could be either still or rotating at a frequency $\nu_B$. In this latter case the frequency $\nu_\psi$ of the effect is $4\nu_{\rm P}-2\nu_B$ or $4\nu_{\rm P}+2\nu_B$ according to whether the magnet is rotating in the same direction as the wave plates or in the opposite direction.

Note that in this configuration the defects of the two wave plates accumulate and cannot be singularly identified. To overcome this difficulty a frequency-doubled green laser is planned to be implemented on the same light path as the infrared one. To green light, in fact, a half-wave plate appears as a full-wave plate, and only the deviations from the nominal retardation will count. By employing the green laser the two wave plates can be rotated at different frequencies thus allowing the characterisation of each optical element.

\begin{figure}[t]
\begin{center}
\includegraphics[width=8.4cm]{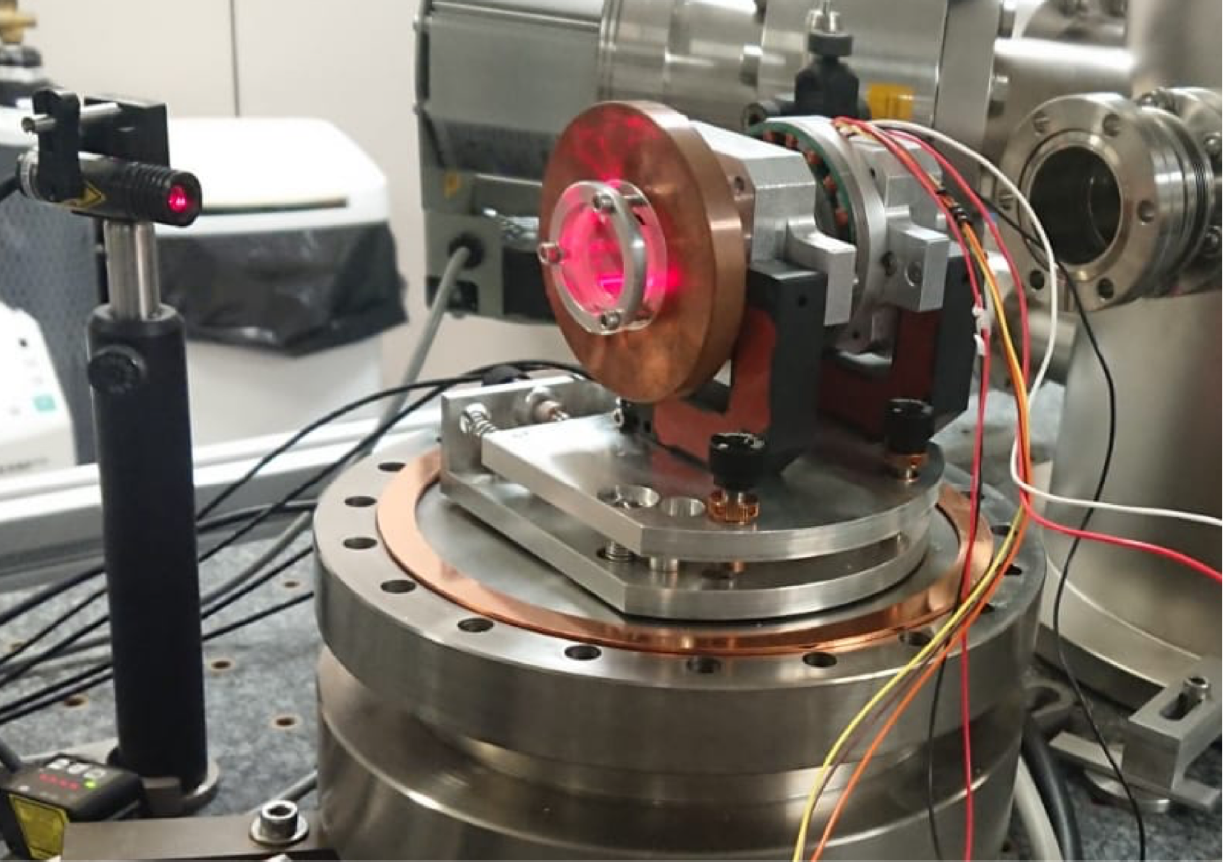}
\end{center}
\caption{A photograph of the rotation stage prototype. The optical element is fastened at the near end of the device.}
\label{Stage}
\end{figure}

The home-made rotation stage employed in the present work is shown in figure~\ref{Stage}; two such devices have been realised. The optical element is held at one end of a hollow cylindrical shaft sustained by two self-aligning precision ball bearings placed at a distance of $\approx5$~cm one from the other. Each ball bearing is fastened to a $200~\mu$m piezoelectric $xy$ linear translation stage (red-black ``L'' shapes in the figure) allowing dynamic fine adjustments of the rotation axis both in translation and in tilt angle; maximum tilt angle is $\pm200~\mu{\rm m}/5~{\rm cm}\approx\pm4$~mrad; the actual response of each of the eight piezo channels has been dynamically calibrated by optical means up to a few tens of hertz. The piezos are supported by a $xy\,\theta_x\theta_y$ positioning table equipped with five independent long-travel M6$\times$0.25 fine adjustment screws (three vertical, two lateral). The table sits on the bottom of a CF150 vacuum chamber.

The shaft of the rotation stage is driven by an 18-pole brushless motor whose windings are visible in the figure: the stator is fastened to the cage of one of the ball bearings whereas the rotor is connected to the shaft. There is enough play between the rotor and the stator for axis orientation. The three phases of the motor are powered by three audio current amplifiers driven by as many identical sine generators at $120^\circ$; in practice, most of the time only two of the three channels were powered. Rotation frequencies of the plates up to $\nu_{\rm P}=15$~Hz have been used. Unlike commercial motor drivers, no adjustment of the phase of the currents relative to the instantaneous phase of the rotor is made. In this way, the system, although able to apply only a light torque on the load, guarantees a long-term phase-locking of the rotation with the waveform of the generators, all locked to a single master clock. The same system was employed by the PVLAS experiment to realise phase-controlled rotation runs lasting longer than $10^6$~s \cite{PVLAS2016,PVLAS2020,Universe}. A 2.7~kg$\cdot$cm$^2$ copper fly-wheel, visible in the picture, helps dampen the angular oscillations of the load associated with the drive system thereby reducing the instantaneous relative phase fluctuations of the two wave plates. Unlike the case of the half-ton magnets, however, these oscillations are not negligible and represent one of the major issues of the set-up, heavily affecting the extinction, as will be discussed below.

The light source is an NPRO 1064~nm Nd-YAG laser from which a power $P_0\approx P_{\rm out}\approx P_\parallel\approx10$~mW is extracted. The polariser and the analyser are high quality Glan-Laser prisms with extinction $\sigma^2\lesssim10^{-7}$. The Photo-Elastic Modulator adds to the beam polarisation a controlled ellipticity $\eta_0\approx10^{-2}$ modulated at $\nu_{\rm m}\approx50$~kHz. The extinguished power $P_\perp$ of equation~\eqref{ExtinguishedBeam} is collected by an InGaAs photodiode with quantum efficiency $q\approx0.7$~A/W and is amplified with a transimpedance $G=10^4$~V/A.

The average relative phase of the two wave plates is easily adjustable, allowing to reach extinction. Moreover, the absolute position of the analyser has been chosen so as to minimise, at extinction, the sum of the two $\alpha_0$ terms of equation~\eqref{DefectSymmetry} for the two wave plates. This operation had to be done only once, as its result remains encoded in the position of the analyser.

The alignment of the $\hat{\mathbf{n}}$ axis is obtained by probing a surface of the rotating plates with a red laser diode (also visible in figure~\ref{Stage}), observing the movements of the $\approx45^\circ$ reflected beam with a position sensitive photodiode and acting on the retaining ring of the optical element which is able to slightly tilt this axis; in this way the wobble of the reflected beam can be reduced to about 1~mm at a distance of $\approx1$~m. The alignment of the laser beam is obtained observing the back-reflection from the surface of the plates and minimising the incidence angle $\theta_i$ by fine adjusting the position of the whole assembly of the rotator. Also in this case a precision of $\approx1$~mm over 1~m was obtained. The two alignments are not accurate enough for the future experiment at CERN, but do guarantee that the spurious signals observed in the context of the present work cannot be attributed to angular misalignments.

\section{Results and discussion}

\subsection{Rotating zero-wave plates}

\begin{figure}[bht]
\begin{center}
\includegraphics[width=8.4cm]{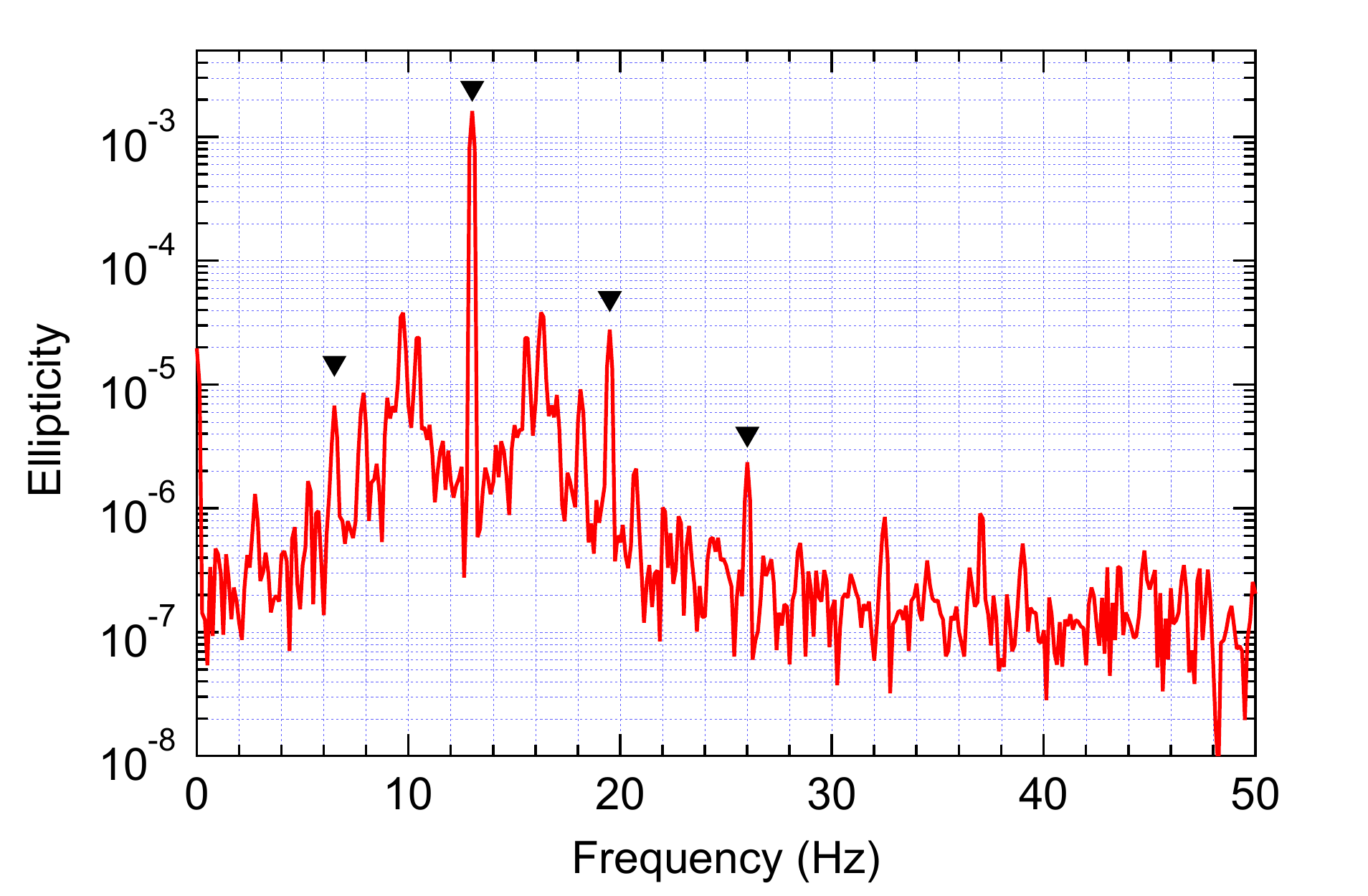}
\end{center}
\caption{A typical demodulated ellipticity spectrum of the optical zero-wave GP rotating at $\nu_{\rm P}=6.5$~Hz. The black triangles mark the first four harmonics of this frequency. A Hanning window and an 8~s integration time have been used; effective time due to the windowing is 5.3~s.}
\label{Rotating0Plate}
\end{figure}

Figure~\ref{Rotating0Plate} shows a typical frequency spectrum of the ellipticity given by equation~\eqref{Psi0}, obtained demodulating the PDE signal at the modulation frequency $\nu_{\rm m}$ of the PEM. The spectrum has been measured in the geometry of figure~\ref{SetUp}a with the GP assembly constituting a zero-wave plate rotating at a frequency $\nu_{\rm P}=6.5$~Hz. The spectrum obtained using the CWP assembly is much the same. All the features of the spectrum are unwanted spurious signals. Nonetheless, whereas the appearance of the first three harmonics was expected, the presence of the fourth harmonic is a serious threat for the proposed polarimetric method. In the following we discuss all the signals and propose a workaround solution to this threat.

\subsubsection{The $2\nu_{\rm P}$ structures}

The spectrum in figure \ref{Rotating0Plate} is dominated by the $\psi_{2\nu_{\rm P}}$ peak at the frequency $2\nu_{\rm P}=13$~Hz, which is due to a small residual static optical path difference of the nominally neutral plate [$\alpha_0$ of equation \eqref{DefectSymmetry} inserted into equation~\eqref{ExtinguishedIntensity}]. From the $\psi_{2\nu_{\rm P}}$ value one deduces the value of this optical path difference
\[
\Delta {\cal D}=\int\Delta n\,dz=\frac{\lambda}{\pi}\,\psi_{2\nu_{\rm P}}\approx0.5~{\rm nm}.
\]
This value approximately holds for both the rotating optical elements employed: for the optical glass GP it corresponds to a birefringence averaged over the 1~cm thickness of the element of $\Delta n_{\rm GP}\approx5\times10^{-8}$, whereas for the CWP assembly it defines an effective thickness error of
\[\Delta D_{\rm CWP}=\frac{\lambda}{\pi\Delta n_{\rm quartz}}\,\psi_{2\nu_{\rm P}}\approx40~{\rm nm}.\]

The value of $\psi_{2\nu_{\rm P}}\approx1.5\times10^{-3}$ corresponds to an equivalent $\alpha\approx3\times10^{-3}$. Given a desired value of the cavity amplification factor $N\approx500$ this results in a total ellipticity $\gtrsim1$ and therefore an undefined polarisation between the Fabry-Perot mirrors. For this reason the amplification with the Fabry-Perot cavity is not included in this paper.

On the two sides of the $2\nu_{\rm P}$ peak, about hundred times smaller than the central peak, two broad bumps can be seen which do not originate from the optics, but instead from the mechanics. As said before, the rotation is driven by a three-phase fixed-frequency sinusoidal current. The angular position of the rotor oscillates chaotically around the nominal value $\phi_{\rm P}(t)=2\pi\nu_{\rm P}t$. The two broad features are generated by the modulation of the ellipticity at $2\nu_{\rm P}$. The oscillation spectrum is determined by the inertia of the load and the stiffness due to the current intensity in the stator. By playing with these two ingredients the position of the bumps can be moved farther or closer to the central peak.

\subsubsection{The $\nu_{\rm P}$ and $3\nu_{\rm P}$ harmonics}

As far as the $\nu_{\rm P}$ and $3\nu_{\rm P}$ peaks are concerned, they are of order $10^{-5}$ as discussed in section \ref{IntrinsicSystematics}. There we assumed the two peaks derived from the beating of the $\nu_{\rm P}$ component of equation~\eqref{LightPath} and the $2\nu_{\rm P}$ dependence of equation~\eqref{ExtinguishedIntensity}; as a consequence, the two side-bands should have the same amplitude, which is clearly not the case. Moreover, the amplitude and the ratio of the two peaks vary with the positioning of the plate. We postulate therefore the existence of at least one more effect synchronised with the rotating plate but not originating from the plate itself. This could be a modulation of the direction of the beam emerging from the rotating plate, coupled with a birefringence map inside the ellipticity modulator. This effect would sum vectorially (in amplitude and phase) with the ellipticity component at $\nu_{\rm P}$, making it different from the $3\nu_{\rm P}$ component. This would explain the different amplitudes of the two peaks and their dependence on the alignment.

\subsubsection{The $4\nu_{\rm P}$ signal}

Let us now discuss the peak at $4\nu_{\rm P}$ in figure~\ref{Rotating0Plate}. We will show that it originates from a transverse oscillation of the rotation axis coupled with the gradient of the optical path difference of the rotating plate: in this condition, the quantity $\alpha_{\nu_{\rm P}}$ in equation~\eqref{DefectSymmetry} acquires a time dependence at the rotation frequency.

\begin{figure}[bht]
\begin{center}
\includegraphics[width=8.4cm]{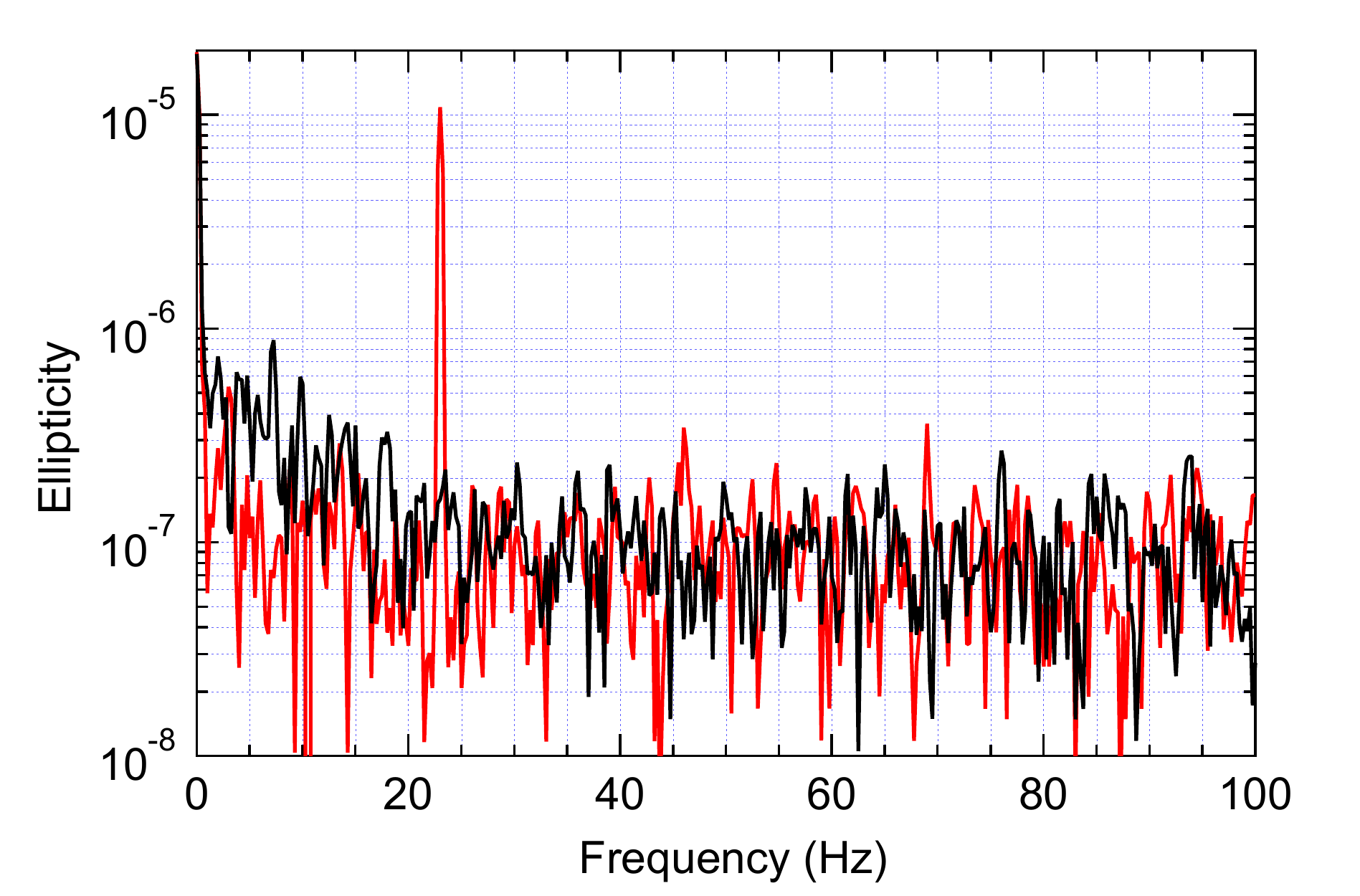}
\end{center}
\caption{Demodulated ellipticity spectra of the non-rotating 1~cm thick optical zero-wave GP oscillating laterally by $\pm30~\mu$m at $\nu_{\rm T}=23$~Hz. Two orthogonal directions of oscillations are shown, one of them giving a minimum response at the modulation frequency.}
\label{Still0Plate}
\end{figure}

Consider the ellipticity spectrum of figure~\ref{Still0Plate} for which the glass plate is non rotating and, instead, is oscillating transversally at $\nu_{\rm T}=23$~Hz with an amplitude $r_{\phi_{1,0}}=30~\mu$m by means of a synchronous displacement of the two parallel piezo stages. In this way the impact point of the laser beam on the surface of the plate moves back and forth. We note explicitly that, as expected, a modulation of the incidence angle $\theta_i$ without modulation of the impact point is found to produce no apparent response in the ellipticity spectra. Two spectra are shown in the figure: the first corresponds to the direction of oscillation which minimises the ellipticity response at the modulation frequency (black) and the second to the orthogonal direction (red) generating an ellipticity amplitude $\psi(\nu_{\rm T}) \approx 10^{-5}$. This latter is hence the direction of the gradient of the optical path difference of equation~\eqref{GradientDeltanD} and one can write
\[
\frac{\pi}{\lambda}\,\vec{\mathbf{r}}_{\phi_{1,0}}\cdot\vec{\mbox{\boldmath$\nabla$}}(\Delta n\,D)=\psi(\nu_{\rm T})\approx10^{-5}
\]
where $\phi_{1,0}$ is the azimuthal phase of the gradient [see equation~\eqref{Phi_s,0}]. Hence
\begin{equation}
\left|\vec{\mbox{\boldmath$\nabla$}}(\Delta n\,D)\right|= \left|\vec{\mbox{\boldmath$\beta$}}\Delta n+D\vec{\mbox{\boldmath$\nabla$}}(\Delta n)\right|\approx10^{-7}.
\label{gradient}
\end{equation}
An identical behaviour and similar spectra have been obtained for the CWP assembly. The interpretation of the spectra is however slightly different in the two cases: for the glass plate, with $\Delta n_{\rm GP}\approx5\times10^{-8}$, the first term of the gradient in equation~\eqref{gradient} is too small to play a role in the observed peak given that $\beta\sim10^{-5}$~rad; from the second term one finds a value for the transverse gradient of the birefringence, averaged along the optical path of the light
\[
|\vec{\mbox{\boldmath$\nabla$}}(\Delta n_{\rm GP})|\approx10^{-5}~{\rm m}^{-1}.
\]
For the CWP assembly, instead, both terms in equation~\eqref{gradient} might contribute: the first term alone would give $\beta\approx10^{-5}$, whereas the second term alone would give $|\vec{\mbox{\boldmath$\nabla$}}(\Delta n_{\rm quartz})|\approx6\times10^{-5}$~m$^{-1}$.

\begin{figure}[bht]
\begin{center}
\includegraphics[width=8.4cm]{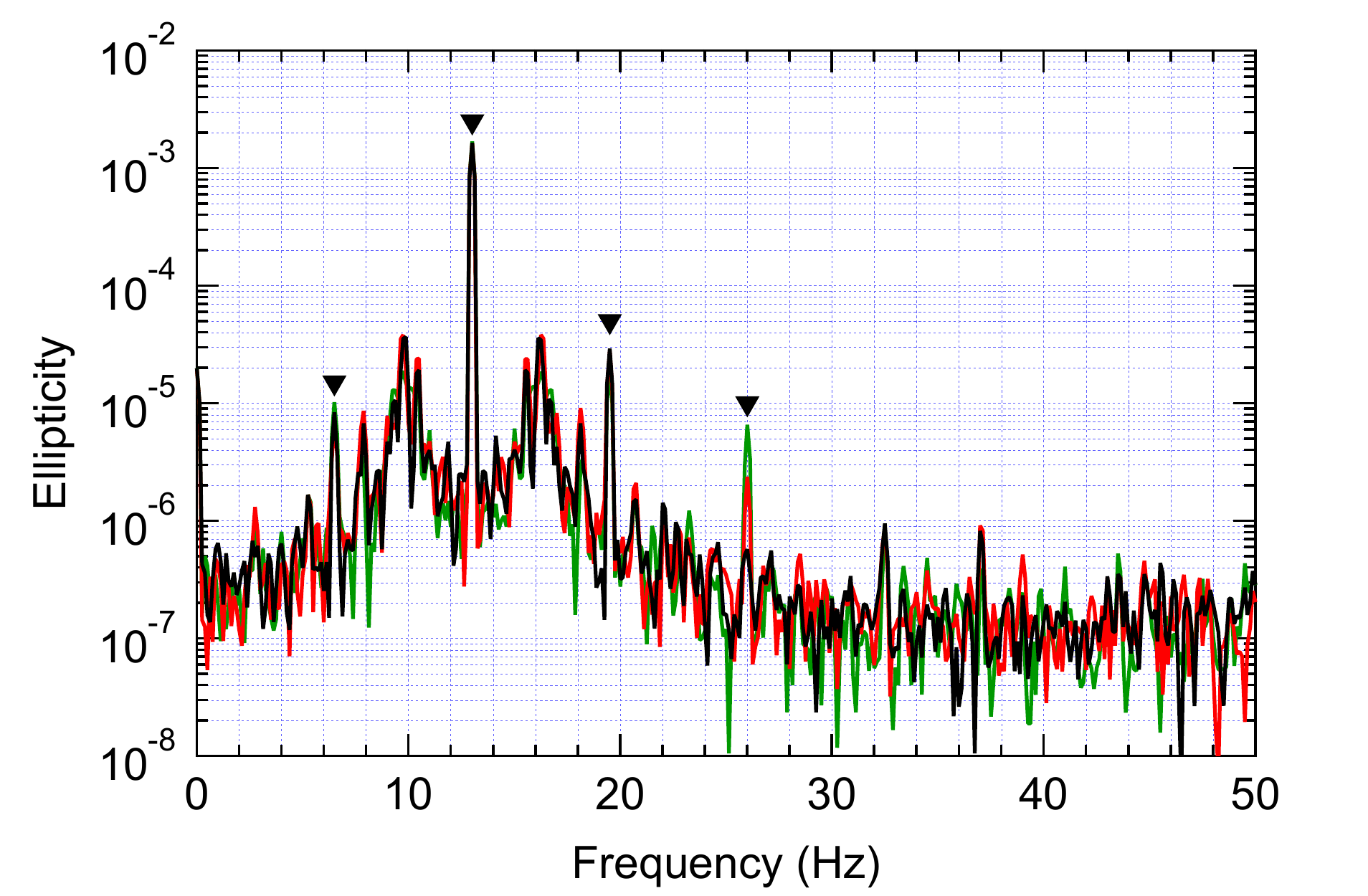}
\end{center}
\caption{Demodulated ellipticity spectra of the zero-wave GP rotating at $\nu_{\rm P}=6.5$~Hz. The black triangles mark the first four harmonics of this frequency. The spectrum with the highest fourth harmonics peak has been recorded while modulating the transverse position of the plate at the frequency $\nu_{\rm P}$ by $\pm30~\mu$m, with the right phase to maximise the response. For the lowest peak spectrum the modulation has the opposite phase and a $15~\mu$m amplitude. The intermediate peak spectrum has no modulation (see figure~\ref{Rotating0Plate}).}
\label{DoubleModulation}
\end{figure}

The combined effect of $\vec{\mbox{\boldmath$\nabla$}}(\Delta n\,D)$ and an axis oscillation during rotation is shown in figure~\ref{DoubleModulation}. In this figure three ellipticity spectra of the rotating zero-wave GP are shown. While the plate rotates, its transverse position is modulated sinusoidally at the frequency of rotation. From the figure it is apparent that this modulation visibly affects only the fourth harmonic. The phase and amplitude of the modulation can be chosen so as to make the peak amplitude larger (green) or to cancel it (black). This behaviour proves what we anticipated: the fourth harmonic peak in figure~\ref{Rotating0Plate} is due to a transverse oscillation of the rotation axis synchronous with the wave plate rotation. We attribute this modulation to $\sim 10\;\mu$m mechanical tolerance of the ball bearings coupled with the transverse gradient of the optical path difference through the optical element. 

\begin{figure}[b]
\begin{center}
\includegraphics[width=8.4cm]{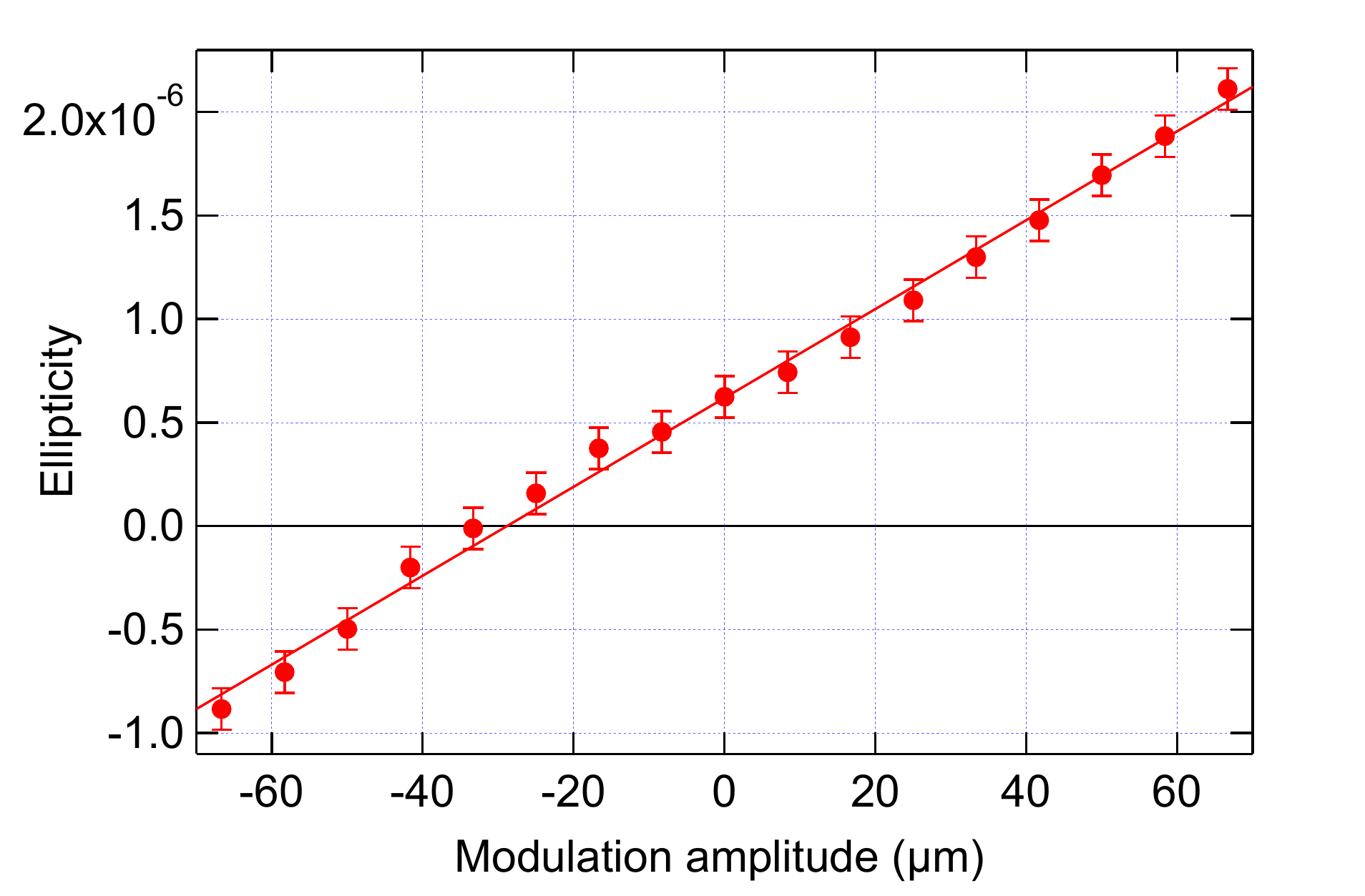}
\end{center}
\caption{Induced ellipticity $4\nu_{\rm P}$ as a function of the modulation amplitude of the rotation axis at $\nu_{\rm P} = 6.5$~Hz (with the phase $\phi_{1,0}$ set for maximum ellipticity) for the zero-wave CWP assembly. Superimposed is a linear fit indicating a slope $\psi_{4\nu_{\rm P}}/r_{\phi_{1,0}} \approx 2\times10^{-8}\;\mu$m$^{-1}$.}
\label{cal}
\end{figure}

In figure~\ref{cal} we report a calibration of the $4\nu_{\rm P}$ ellipticity component generated by the axis transverse modulation at the rotation frequency as a function of amplitude, using the CWP assembly. The phase of the modulation is chosen to be the phase $\phi_{1,0}$ of the wedge so as to maximise the induced ellipticity. A linear fit indicates a slope of $\psi_{4\nu_{\rm P}}/r_{\phi_{1,0}} \approx 2\times10^{-8}\;\mu$m$^{-1}$. The same calculations on the data of figure~\ref{DoubleModulation}, considered linear, would give $\psi_{4\nu_{\rm P}}/r_{\phi_{1,0}} \approx10^{-7}\;\mu$m$^{-1}$. Therefore considering the value to be measured in the expected conditions at CERN, reported in expression \eqref{ell_cern}, a requirement for the transverse oscillation of the rotation axis ensues:
\begin{equation}
    r^{\rm (CERN)}_{\phi_{\nu_{\rm P}}}\lesssim10^{-12}\;{\rm m},
\end{equation}
an unreasonable value to control. The appearance of this spurious signal represents a serious threat for the possibility of measuring very small birefringences with the method proposed in Ref.~\cite{RotatingWavePlates}. In the next section, where birefringence measurements are presented, a possible workaround will be described.

We note explicitly that the same fourth harmonic could be associated to the term $\alpha_{2\nu_{\rm P}}$ of equation~\eqref{DefectSymmetry}. However, unlike the first order term, the second and higher order terms imply the existence of a center of symmetry we never observed. Moreover, higher order terms would not be able to produce the behaviour shown in figure~\ref{DoubleModulation}.

\subsection{Two rotating half-wave plates}

In this section we present the birefringence measurements taken with the polarimeter of figure~\ref{SetUp}b. The new polarimetric scheme has been tested with the Cotton-Mouton effect \cite{Rizzo1997}. This effect is analogous to the vacuum magnetic birefringence described by equation~\eqref{VMB}, but is far more intense already at low gas pressures. The birefringence generated in a gas at pressure $P$ by a magnetic field $B_{\rm ext}$ is given by the expression
\[
\Delta n_{\rm CM}=\Delta n_{\rm u}B_{\rm ext}^2P
\]
where $\Delta n_{\rm u}$ is a unitary birefringence usually expressed in tesla$^{-2}$atmosphere$^{-1}$.

\subsubsection{Measurements with a static magnetic field}

\begin{figure}[bht]
\begin{center}
\includegraphics[width=8.4cm]{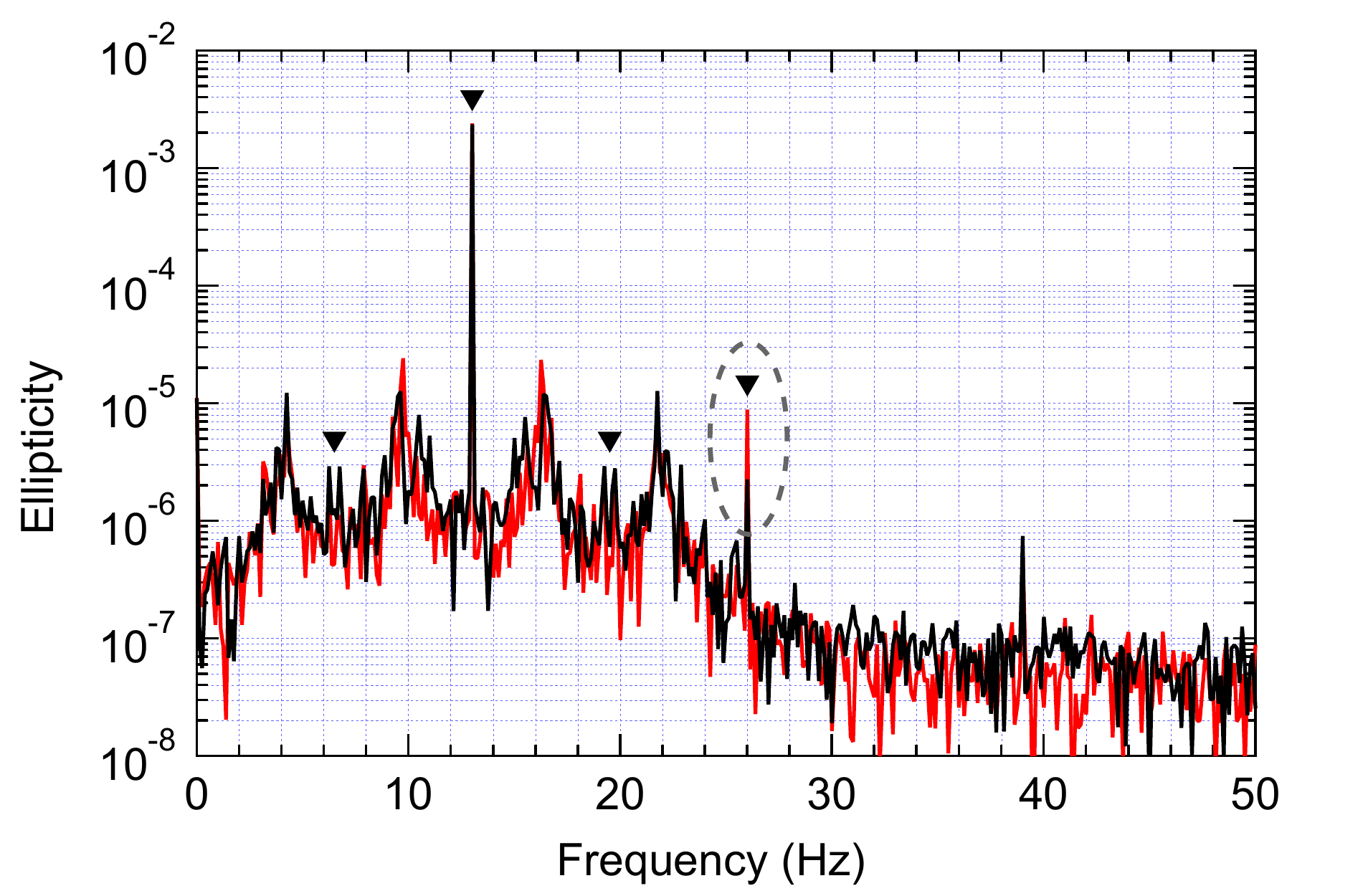}
\end{center}
\caption{Cotton-Mouton ellipticity spectrum of air obtained with the rotating wave plates polarimeter. The two spectra have been recorded with the magnetic fields of the two stationary dipole magnets oriented parallel (red) and orthogonal (black) to each other. A uniform window and 10 vector averages lasting 8~s each have been used.}
\label{CottonMouton}
\end{figure}

In figure~\ref{CottonMouton} two ellipticity spectra of air at atmospheric pressure are shown, taken with the two PVLAS dipole magnets kept stationary and oriented either parallel or orthogonal to each other. In this second case the ellipticity acquired by the polarisation inside one magnet is canceled by the effect of the other, leaving at the frequency $4\nu_{\rm P}$ only the effect of the systematic. The vector difference (in amplitude and phase) of the two $4\nu_{\rm P}$ signals is thus a measurement of the Cotton-Mouton effect in air at 1~atm. The value obtained is compatible with the known values of the Cotton-Mouton constants of Nitrogen and Oxygen and their stoichiometry. We conclude that the method of Ref.~\cite{RotatingWavePlates} works in principle. However, the coincidence in frequency of the Cotton-Mouton signal with the above mentioned systematic effect makes the measurement of figure~\ref{CottonMouton} viable only for large signals, but certainly not for the vacuum magnetic birefringence.

\begin{figure}[bht]
\begin{center}
\includegraphics[width=8.4cm]{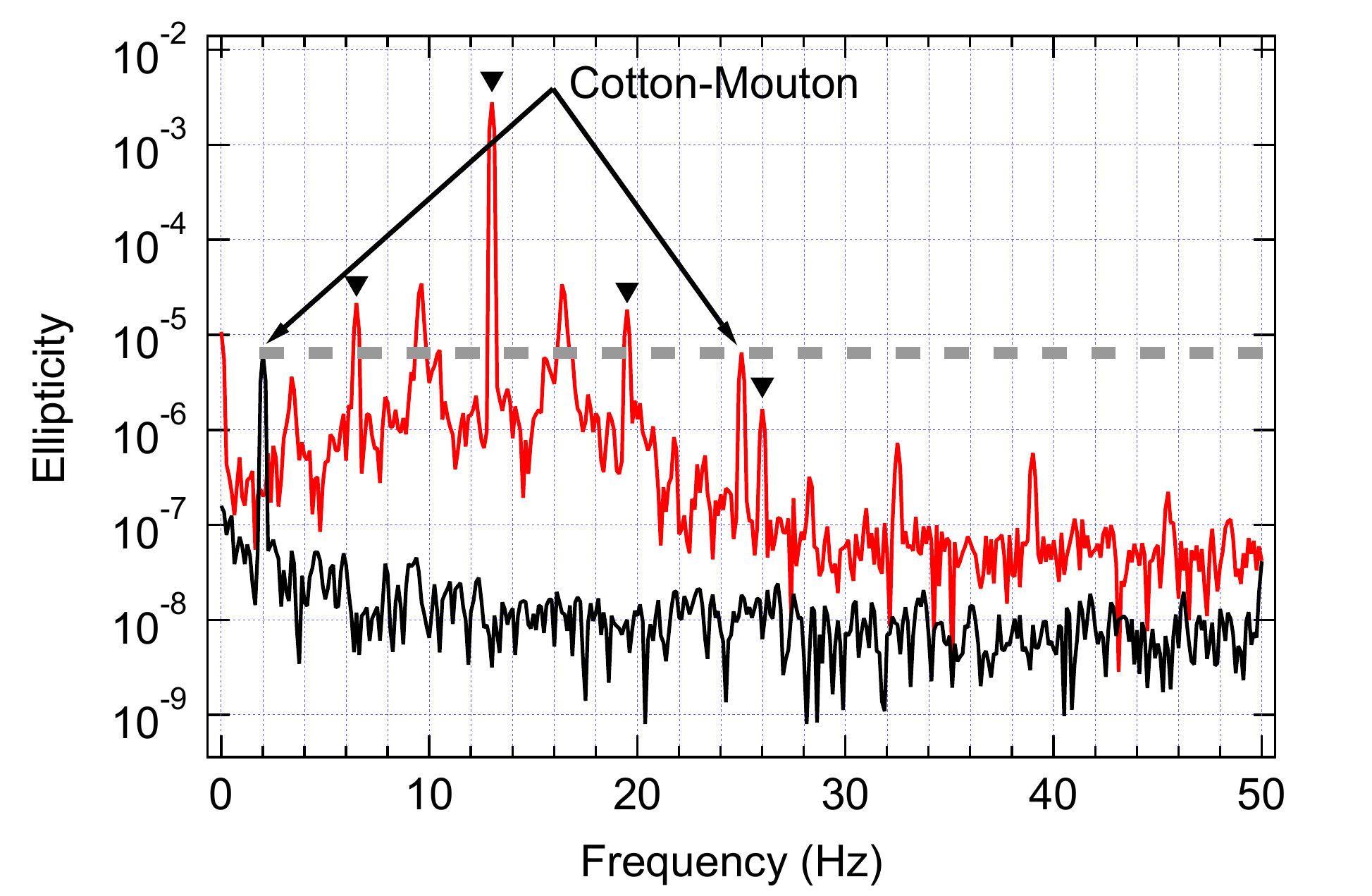}
\end{center}
\caption{Cotton-Mouton ellipticity spectra of air at atmospheric pressure. The two spectra have been recorded with one dipole magnet rotating. For the red spectrum the wave plates also rotate at $\nu_{\rm P}=6.5$~Hz. The black triangles mark the first four harmonics of this frequency. A Hanning window and 10 vector averages lasting 8~s each have been used; effective time due to the windowing is 5.3~s/average.}
\label{ModulatedField}
\end{figure}

\subsubsection{Measurements with a slowly modulated magnetic field}

A possible workaround for the systematic described above is shown in figure~\ref{ModulatedField} where two Cotton-Mouton ellipticity spectra are compared both obtained with one of the PVLAS dipole magnets in rotation. In the first case (red plot) the half-wave plates are rotating at $\nu_{\rm P} = 6.5$~Hz with the PVLAS magnet rotating at $\nu_B^{(\rm red)}=0.5$~Hz in the same direction. In this case the Cotton-Mouton effect is found at the frequency $\nu_\psi=4\nu_{\rm P}-2\nu^{(\rm red)}_B=25$~Hz. In the second case (black plot) the half-wave plates were stationary and the PVLAS magnet was rotating at $\nu_B^{(\rm black)}=1.0$~Hz. In this second case the only signal in the spectrum is the Cotton-Mouton effect at $\nu_\psi=2\nu^{(\rm black)}_B=2$~Hz. Note that the amplitude of the Cotton-Mouton signals is the same in the two spectra, indicating that a modulation of the magnetic field effectively separates the magnetic birefringence from the spurious signal. Also notice the difference in the noise levels in the flat regions. This is due to the relative angular fluctuations of the two rotating half-wave plates and will be discussed in detail in Section \ref{Rotation measruements}.

This new scheme (namely the one with slowly rotating magnets) was further tested by measuring the Cotton-Mouton effect in pure Nitrogen gas at $T=296\pm1$~K. The measurements were preceded by an absolute calibration of the polarimeter by measuring the polarisation rotation signals (hence with the quarter-wave plate of figure \ref{SetUp}b inserted) as a function of different input polarisation directions, measured by an encoder having about $1~\mu$rad resolution. The data were taken by spanning an input polarisation direction range of 8~mrad. The measured values obtained by applying equation~\eqref{Psi0} were fitted with a linear function, resulting in a $0.9718\pm0.0024$ slope with a $\chi^2/{\rm d.o.f.}=21.3/19$. We used this slope value as a normalisation factor for the polarimetric measurements; the slope differs from unity for several reasons: uncertainty on the absolute $P_{\rm out}$ power, incomplete light collection on the photodiode, uncertainty on the photodiode quantum efficiency, differences in the calibrations of the $\nu_{\rm m}$ and $2\nu_{\rm m}$ lock-in amplifiers, etc.

\begin{figure}[bht]
\begin{center}
\includegraphics[width=8.4cm]{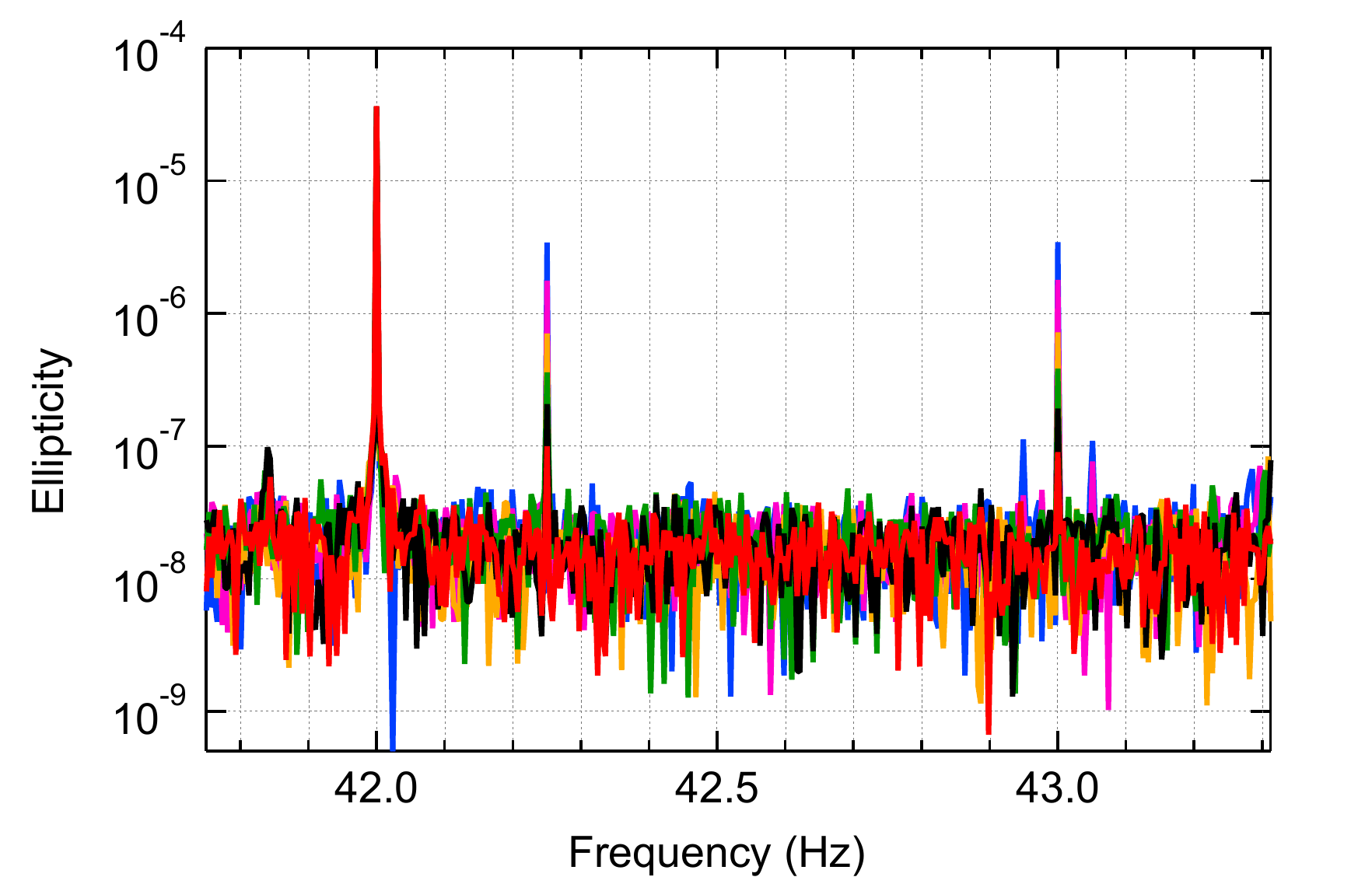}
\end{center}
\caption{Ellipticity spectra of the magnetic birefringence of six different pressures of Nitrogen gas taken with the wave plates rotating at $\nu_{\rm P}=10.5$~Hz and the magnets rotating at $\nu_{B\alpha}=0.125$~Hz and $\nu_{B\beta}=0.5$~Hz.} 
\label{N2Spectra}
\end{figure}

In the Nitrogen Cotton-Mouton measurements, the two wave plates were rotating at $\nu_{\rm P}=10.5$~Hz and the two magnets were rotating in the opposite direction with respect to the wave plates at the frequencies $\nu_{B\alpha}=0.125$~Hz and $\nu_{B\beta}=0.5$~Hz. This generates an ellipticity spectrum having peaks at $4\nu_{\rm P}=42$~Hz, $4\nu_{\rm P}+2\nu_{B\alpha}=42.25$~Hz and $4\nu_{\rm P}+2\nu_{B\beta}=43$~Hz. The measurements taken at the pressures 25.0, 51.7, 100, 200, 500 and 1000 mbar can be seen in figure~\ref{N2Spectra}, where the spectra of the different pressures are superimposed with different colors. It is apparent that the $4\nu_{\rm P}$ peak does not depend on pressure, whereas the signals at $4\nu_{\rm P}+2\nu_{B\alpha}$ and $4\nu_{\rm P}+2\nu_{B\beta}$ are identical to each other and scale with pressure. Around the 43~Hz peak two side bands can be seen at the higher pressures due to a slow oscillation of the phase of the faster rotating magnet around the central value $\phi_{B\beta}(t)=\nu_{B\beta}\,t$, due to the driving technique. The standard deviation of the integrated noise floor is $\sigma_\psi = 1.3\times10^{-8}.$

\begin{figure}[t]
\begin{center}
\includegraphics[width=8.4cm]{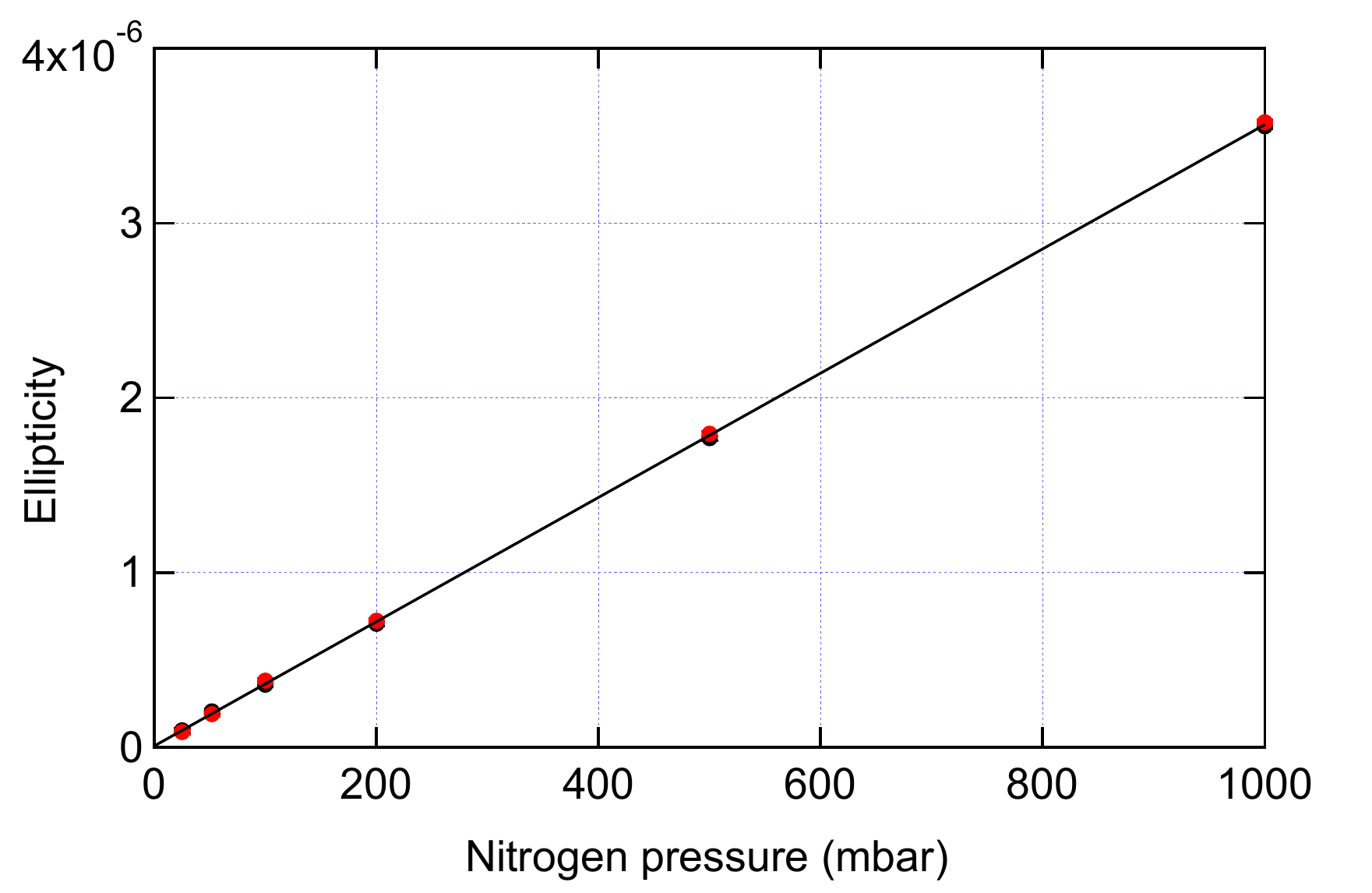}
\end{center}
\caption{Global linear fit of the $4\nu_{\rm P}+2\nu_{B\alpha}$ and $4\nu_{\rm P}+2\nu_{B\beta}$ ellipticity data of figure~\ref{N2Spectra} as a function of the gas pressure. Fit parameters are: $(3.555\pm0.010)\times10^{-9}$/mbar slope, $(10\pm5)\times10^{-9}$ intercept and $\chi^2/{\rm d.o.f} = 9.4/10$. The statistical $1\sigma$ uncertainty for each data point is $\sigma_\psi=1.3\times10^{-8}$.} 
\label{N2CM}
\end{figure}

Figure~\ref{N2CM} shows the global linear fit of the amplitudes of the $4\nu_{\rm P}+2\nu_{B\alpha}$ and $4\nu_{\rm P}+2\nu_{B\beta}$ signals as a function of the gas pressure resulting in $(3.555\pm0.010)\times10^{-9}$/mbar slope, $(10\pm5)\times10^{-9}$ intercept and $\chi^2/{\rm d.o.f} = 0.94$ (statistical). All the points have been fitted by a single line. The slope of the line corresponds to
\[
\Delta n_{\rm u}=(2.380\pm0.007^{\rm (stat)}\pm0.024^{\rm (sys)})\times10^{-13}~{\rm T}^{-2}{\rm atm}^{-1}
\]
which is the most precise measurement of the unitary birefringence of Nitrogen \cite{Rizzo1997,Chen2007,Mei2009,CPL2014}. The systematic uncertainty dominates and is estimated as
\[
\sigma^{\rm(sys)}=\sqrt{\left[\frac{2\Delta T}{T}\right]^2+\left[\frac{\Delta(B_{\rm ext}^2L_B)}{B_{\rm ext}^2L_B}\right]^2}
\]
where $T$ is the absolute Nitrogen gas temperature and $\Delta n_{\rm u}\propto T^{-2}$. Note that the fit gives a small $2\sigma$ intercept when one would have expected it to be zero. At present we do not have an explanation for it, but setting it to zero changes the slope only by 4\permil.

\subsubsection{Line shape of the spurious signal at $4\nu_{\rm P}$}
\label{Induttanza}

\begin{figure}[bht]
\begin{center}
\includegraphics[width=8.4cm]{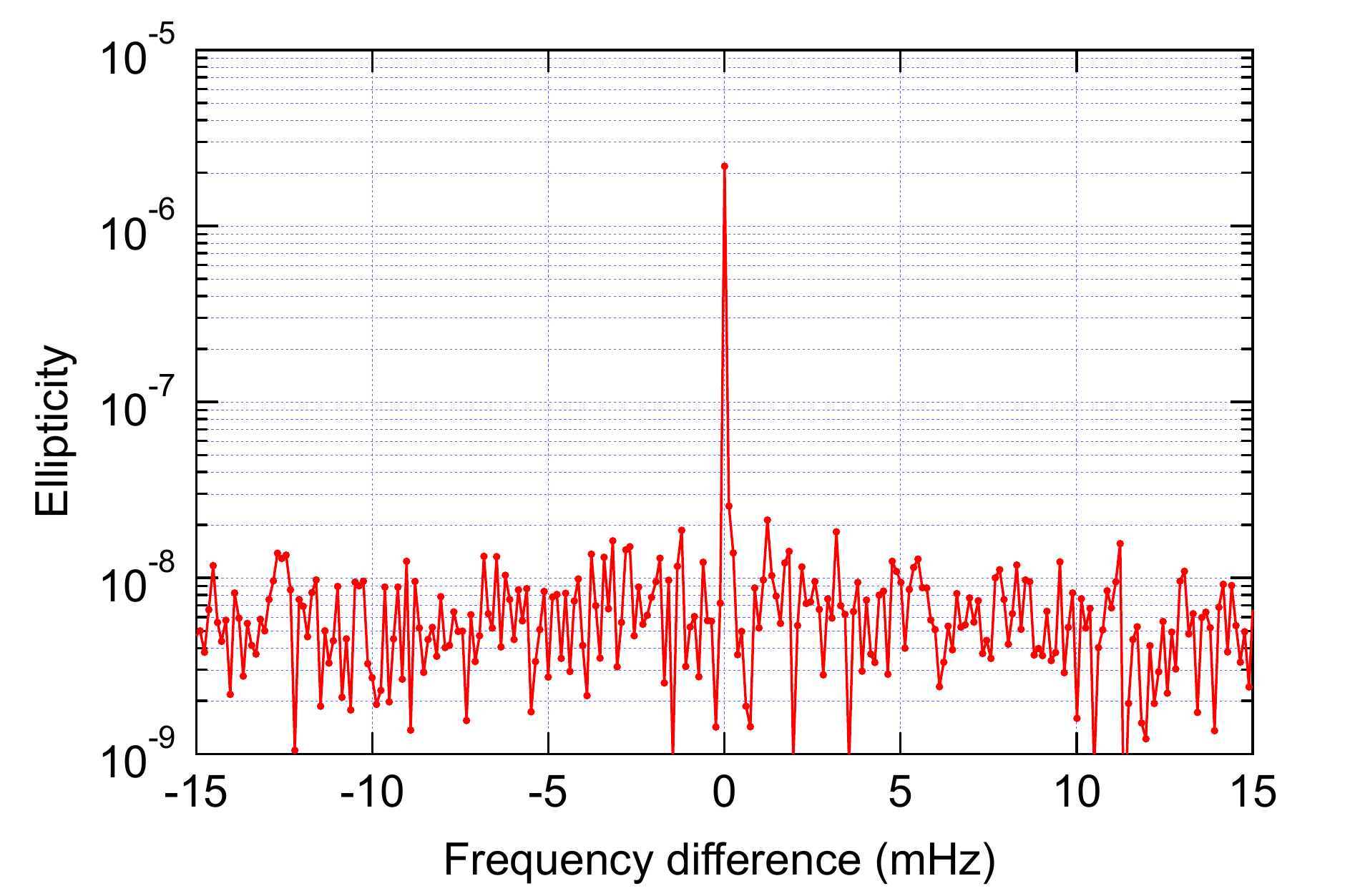}
\end{center}
\caption{Spurious ellipticity signal at $4\nu_{\rm P}=42$~Hz.} 
\label{Width}
\end{figure}

To consider the modulated field strategy as a possibility for measuring the vacuum magnetic birefringence induced by an LHC superconducting dipole magnet one has to guarantee that the noise at the frequency where the ellipticity signal due to the field modulation appears as a side band of the spurious $4\nu_{\rm P}$ peak is dominated by the shot noise. One is led then to study the line shape of the $4\nu_{\rm P}$ spurious signal. Unlike the PVLAS magnets, an LHC dipole cannot be easily modulated. The simplest way is to ramp the current up and down, which would generate two symmetrical peaks around the spurious one. Given the inductance of an LHC dipole of ${\cal L}_{\rm LHC}\approx 100$~mH, a maximum current of $I_{\rm LHC} = 13,000$~A and a maximum power supply voltage of about $V_{\rm max} = 60$~V \cite{Pugnat_private} the maximum modulation frequency will be about 7~mHz. If the tails of the spurious signal extend beyond this value the sensitivity will be compromised. A conclusion on this issue will only be possible at CERN with the complete polarimeter. A preliminary measurement has been done with the polarimeter of figure~\ref{SetUp}b (hence without the Fabry-Perot) showing a SNR of $\approx 300$ without apparent structures above 0.12~mHz. This result is shown in figure~\ref{Width}.

\subsubsection{Rotation measurements}
\label{Rotation measruements}

We note that the noise level in the apparatus is at the moment high and far from shot noise. In fact the dynamical angular position errors of the two stages limit the extinction and hence the sensitivity of the apparatus. A measured value for the extinction coefficient is $P_\perp/P_{\rm out}\approx3.5\times10^{-3}$, dominant over the intrinsic extinction ratio of the polarisers $\sigma^2 \lesssim 10^{-7}$ and over $\eta_0^2/2 \lesssim10^{-4}$. Given the Relative Intensity Noise of the laser at the $\nu_{\rm m}$ modulation frequency ${\rm RIN}(\nu_{\rm m})=10^{-6}/\sqrt{\rm Hz}$ and $\eta_0\approx10^{-2}$, the expected sensitivity above $\approx30$~Hz is \cite{PVLAS2020} \[
S_\psi\approx{\rm RIN}(\nu_{\rm m})\frac{P_\perp/P_{\rm out}}{\eta_0} = 3.5\times10^{-7}/\sqrt{\rm Hz}
\] in agreement with the integrated values in figure~\ref{ModulatedField}, red spectrum, of about $S_\psi/\sqrt{53~{\rm s}}\approx 5\times10^{-8}$. From this figure it is also apparent that the noise of the spectrum obtained with the rotating wave plates is almost an order of magnitude higher than the spectrum with still wave plates.

\begin{figure}[t]
\begin{center}
\includegraphics[width=8.4cm]{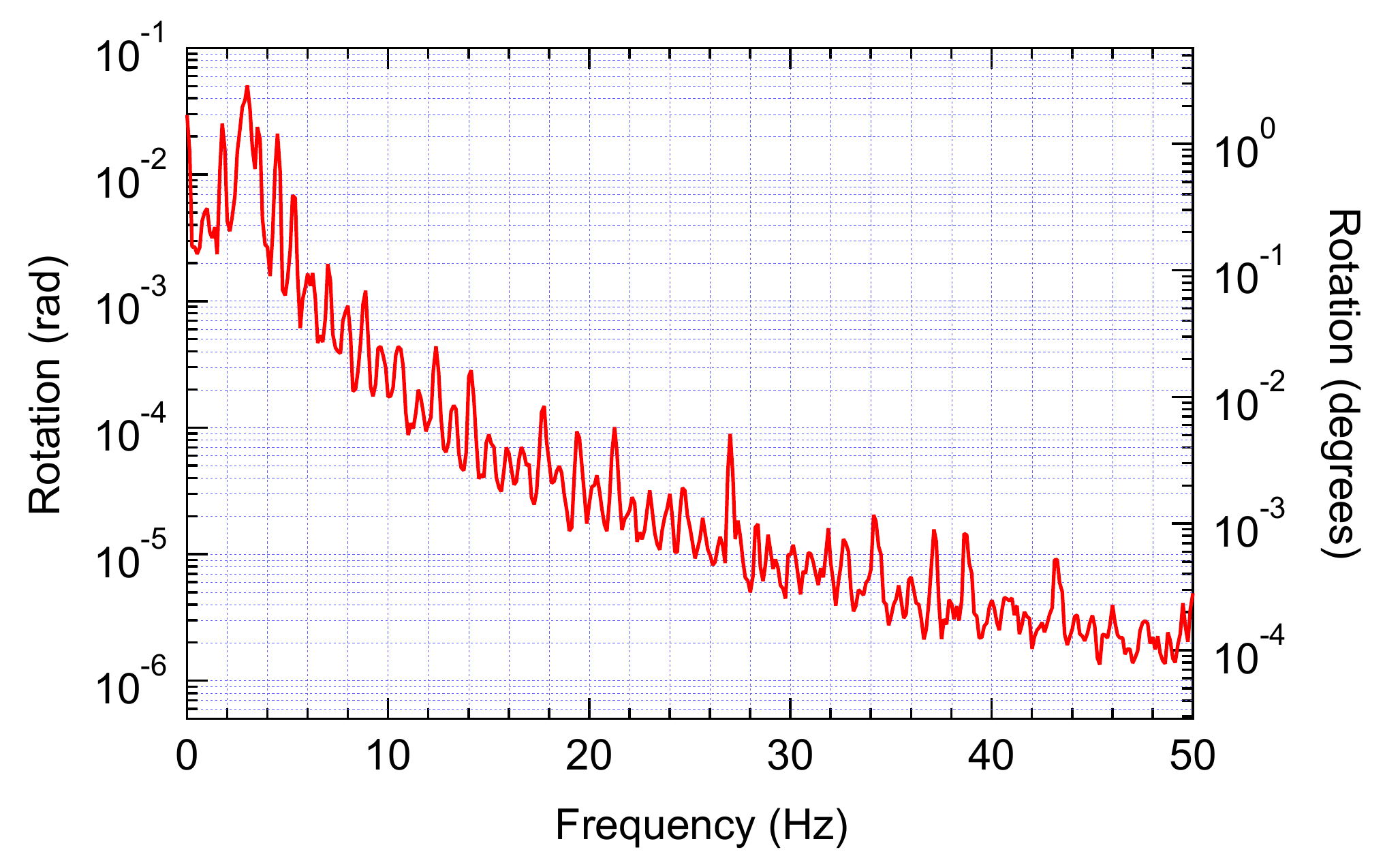}
\end{center}
\caption{Demodulated rotation spectrum obtained with the quarter-wave plate inserted in the scheme of figure~\ref{SetUp}b and the half-wave plates rotating at $\nu_{\rm P}=6.5$~Hz. Integration time 8~s, Hanning window. The rms rotation value is $\approx5^\circ$.} 
\label{RotationSpectrum}
\end{figure}

In figure~\ref{RotationSpectrum} a rotation spectrum is presented measured with the polarimeter of figure~\ref{SetUp}b. A pronounced broad feature peaking at about 0.05~rad is observed at a frequency $\nu\approx3$~Hz due to the angular oscillations of the two rotating wave plates. It is apparent that this feature reproduces the two broad side-bands appearing in Figure~\ref{DoubleModulation}. In fact, as equation~\eqref{ExtinguishedIntensity} shows, if the phases $\phi_1$ and $\phi_2$ of the two wave plates are not constant, there will be a rotation noise in the polarisation. This rotation noise manifests itself as side bands structures of the $2\nu_{\rm P}$ peak in all of the ellipticity spectra shown before, both of the zero-wave plate and of the two half-wave plates, with the structures depending on the rotational inertia and on the current intensity in the motor windings. The resulting rms value of the rotation noise $S_\varphi$ corresponds to $\Delta\varphi_{\rm rms}=\sqrt{\int |S_\varphi|^2\;d\nu}\approx5^\circ$ in agreement with the measured extinction coefficient: $\Delta\varphi_{\rm rms}^2/2\approx P_\perp/P_{\rm out}$. 

\section{Conclusions}

We have tested a new polarimetric method we intend to use in an apparatus to measure the vacuum magnetic birefringence by employing the quasi static field generated by a LHC spare steering magnet. The method is based on the 1979 design by Iacopini and Zavattini \cite{Iacopini1979} and modulates the ellipticity signal by using two co-rotating half-wave plates we plan to place inside a Fabry-Perot cavity \cite{RotatingWavePlates}.

The present study has put in evidence an unbeatable spurious signal which imposes a modification of the original scheme in order to measure the vacuum magnetic birefringence. A workaround has been devised and tested: the birefringence effect should be modulated by slowly varying the magnetic field. We have verified that in this way the spurious and the birefringence signals are separated. The method has been validated with a high precision measurement of the Cotton-Mouton effect in Nitrogen gas. In the case of the superconducting LHC dipole the modulation should be obtained by slowly ramping the current.

A few issues remain to be clarified. If the current of a LHC magnet can be ramped at a frequency $\nu_M\approx5$~mHz, one has to verify that the line shape of the spurious signal has no structures beyond $\nu_M$ down to the level of the vacuum birefringence. Since the line shape depends on the design, the implementation and the mechanical stability of the rotation stages, an improved rotation stage should result in a smaller spurious signal. The polarimeter described in this paper also suffers from the synchronisation errors of the two rotation stages. In fact, the two stages rotate synchronously only on average, whereas the rms error has been measured to be $\approx5^\circ$. The extinguished power deviates from zero with marked oscillations whose effective value correspond to an extinction coefficient $P_\perp/P_{\rm out}\approx3.5\times10^{-3}$. In this condition the intensity noise of the laser largely prevails on shot noise. Moreover, the possibility of rotation measurements is heavily impaired. 

To finalise the VMB@CERN apparatus we will have to implement the Fabry-Perot cavity. To this end it is indispensable to control the phase retardation error of the wave plates. This will be done by regulating the temperatures of the wave plates.


\vspace{2mm}
\section*{Acknowledgments}

We gratefully acknowledge the invaluable technical help of L. Stiaccini, of University of Siena, Dipartimento di Scienze Fisiche, della Terra e dell'Ambiente and of M.~Cavallina, of INFN, Sezione di Ferrara.

This research was partly supported by the MEYS of Czech Republic, Grant No. LM2015058 and by the Leverhulme Trust in the UK, Grant No. RPG 2019-002. One of the authors (GM) acknowledges funding from the European Union's Horizon 2020 research and innovation programme under the Marie Sk{\l}odowska-Curie Grant No. 754496.


\begin{thebibliography}{00}

\bibitem{Dirac1928}
P.A.M. Dirac, Proc. Roy. Soc. London A {\bf117} (1928) 610.
https://doi.org/ \!\!10.1098/rspa.1928.0023
\bibitem{Dirac1931}
P.A.M. Dirac, Proc. Roy. Soc. London A {\bf133} (1931) 60.
https://doi.org/ \!\!10.1098/rspa.1931.0130
\bibitem{Dirac1930}
P.A.M. Dirac, Proc. Roy. Soc. London A {\bf126} (1930) 360.
https://doi.org/ \!\!10.1098/rspa.1930.0013
\bibitem{Heisenberg1936}
W. Heisenberg, H. Euler, Z. Phys. {\bf98} (1936) 714.
https://doi.org/ \!\!10.1007/BF01343663, English translation by W. Korolevski and H. Kleinert, arXiv:physics/0605038
\bibitem{Weisskopf1936}
V. Weisskopf, K. Dan. Vidensk. Selsk., Mat. Fys. Medd. {\bf14} (6) (1936), English translation by D.H. Delphenich.
https://neo-classical-physics.info/ \!\!uploads/3/4/3/6/34363841/weisskopf\_-\_electrodynamics.pdf
\bibitem{Halpern1933}
O. Halpern, 
Phys. Rev. {\bf44} (1933) 855.
https://doi.org/ \!\!10.1103/PhysRev.44.855.2
\bibitem{Euler1935}
H. Euler, B. Kockel, Naturwissenschaften {\bf23} (1935) 246.
https://doi.org/ \!\!10.1007/BF01493898, English translation by D.H. Delphenich.
http://www.neo-classicalphysics.info/ \!\!uploads/3/4/3/6/34363841/euler-koeckel\_-\_scattering\_of\_light\_by\_light.pdf
\bibitem{Euler1936}
H. Euler, Ann. Phys. {\bf26} (1936) 398.
https://doi.org/ \!\!10.1002/andp.19364180503, English translation by D.H. Delphenich.
http://www.neo-classicalphysics.info/ \!\!uploads/3/4/3/6/34363841/euler\_-\_scattering\_of\_light\_by\_light.pdf
\bibitem{Karplus1950}
R. Karplus, M. Neuman, 
Phys. Rev. {\bf80} (1950) 380.
https://doi.org/ \!\!10.1103/PhysRev.80.380
\bibitem{Schwinger1951}
J.S. Schwinger, 
Phys. Rev. {\bf82} (1951) 664.
https://doi.org/ \!\!10.1103/PhysRev.82.664
\bibitem{Kinoshita1990}
T. Kinoshita (Ed.), {\em Quantum Electrodynamics}, Advanced Series on Directions in High Energy Physics Vol. 7 (World Scientific, Singapore, 1990).
\bibitem{Karshenboim2005}
S.G. Karshenboim, Phys. Rep. {\bf422} (2005) 1.
https://doi.org/ \!\!10.1016/j.physrep.2005.08.008
\bibitem{ATLAS}
G. Aad {\em et al.} (ATLAS Collaboration), 
Phys. Rev. Lett. {\bf123} (2019) 052001. https://doi.org/10.1103/PhysRevLett.123.052001
\bibitem{CMS}
A.M. Sirunyan {\em et al.} (CMS Collaboration), 
Phys. Lett. B {\bf797} (2019) 134826. https://doi.org/10.1016/j.physletb.2019.134826
\bibitem{Moulin}
F. Moulin, D. Bernard, Opt. Comm. {\bf164} (1999) 137.
https://doi.org/10.1016/S0030-4018(99)00169-8
\bibitem{Sarazin}
S. Robertson {\em et al.}, Phys. Rev. A {\bf103} (2021) 023524.
https://doi.org/10.1103/PhysRevA.103.023524
\bibitem{DiPiazza}
A. Di Piazza {\em et al.}, Rev. Mod. Phys. {\bf84} (2012) 1177.
https://doi.org/10.1103/RevModPhys.84.1177
\bibitem{Karbstein}
F. Karbstein, Particles {\bf3} (2019) 39.
https://doi.org/10.3390/particles3010005
\bibitem{Mignani}
R.P. Mignani {\em et al.}, 
Mon. Not. R. Astron. Soc. {\bf465} (2016) 492. https://doi.org/10.1093/mnras/stw2798
\bibitem{Toll1952}
J.S. Toll, ``The dispersion relation for light and its applications to problems involving electron pairs'' (Ph.D. thesis), Princeton University, Princeton, NJ, 1952, unpublished.
\bibitem{Erber1961}
T. Erber, 
Nature {\bf190} (1961) 25.
https://doi.org/ \!\!10.1038/190025a0
\bibitem{Klein1964}
J.J. Klein, B.P. Nigam, 
Phys. Rev. {\bf135} (1964) B1279.
https://doi.org/ \!\!10.1103/PhysRev.135.B1279
\bibitem{Baier1967}
R. Baier, P. Breitenlohner, 
Nuovo Cimento B {\bf47} (1967) 117.
https://doi.org/ \!\!10.1007/BF02712312
\bibitem{Bialynicka1970}
Z. Bialynicka-Birula, I. Bialynicki-Birula, 
Phys. Rev. D {\bf2} (1970) 2341.
https://doi.org/ \!\!10.1103/PhysRevD.2.2341
\bibitem{Adler1971}
S.L. Adler, 
Ann. Phys., NY {\bf67} (1971) 599.
https://doi.org/ \!\!10.1016/0003-4916(71)90154-0
\bibitem{GrassiStrini1975}
A.M. Grassi Strini, G. Strini, G. Tagliaferri, 
Phys. Rev. D {\bf19} (1979) 2330.
https://doi.org/ \!\!10.1103/PhysRevD.19.2330
\bibitem{Rizzo1997}
C. Rizzo, A. Rizzo, D.M. Bishop, 
Int. Rev. Phys. Chem. {\bf16} (1997) 81.
https://doi.org/ \!\!10.1080/014423597230316
\bibitem{Maiani1986}
L. Maiani, R. Petronzio, E. Zavattini, 
Phys. Lett. B {\bf175} (1986) 359. https://doi.org/10.1016/0370-2693(86)90869-5
\bibitem{Gies2006}
H. Gies, J. Jaeckel, A. Ringwald, 
Phys. Rev. Lett. {\bf97} (2006) 140402. https://doi.org/10.1103/PhysRevLett.97.140402
\bibitem{PVLAS2016}
F. Della Valle {\em et al.} (PVLAS Collaboration), 
Eur. Phys. J. C {\bf76} (2016) 24.
https://doi.org/ \!\!10.1140/epjc/s10052-015-3869-8
\bibitem{PVLAS2020}
A. Ejlli {\em et al.} (PVLAS Collaboration), Phys. Rep. {\bf871} (2020) 1.
https://doi.org/ \!\!10.1016/j.physrep.2020.06.001
\bibitem{Universe}
G. Zavattini, F. Della Valle, Universe {\bf7} (2021) 252.
https://doi.org/10.1016/10.3390/universe7070252
\bibitem{Iacopini1979}
E. Iacopini, E. Zavattini, 
Phys. Lett. B {\bf85} (1979) 151.
https://doi.org/ \!\!10.1016/0370-2693(79)90797-4
\bibitem{Iacopini1981}
E. Iacopini {\em et al.}, 
Nuovo Cimento B {\bf61} (1981) 21.
https://doi.org/ \!\!10.1007/BF02721700
\bibitem{BFRT1993}
R. Cameron {\em et al.}, 
Phys. Rev. D {\bf47} (1993) 3707.
https://doi.org/ \!\!10.1103/PhysRevD.47.3707
\bibitem{PVLAS1998}
D. Bakalov {\em et al.} (PVLAS Collaboration), 
Quantum Semiclass. Opt. {\bf10} (1998) 239.
https://doi.org/ \!\!10.1088/1355-5111/10/1/027D
\bibitem{PVLAS2008a}
E. Zavattini {\em et al.} (PVLAS Collaboration), 
Phys. Rev. D {\bf77} (2008) 032006.
https://doi.org/ \!\!10.1103/PhysRevD.77.032006
\bibitem{PVLAS2008b}
M. Bregant {\em et al.} (PVLAS Collaboration), 
Phys. Rev. D {\bf78} (2008) 032006.
https://doi.org/ \!\!10.1103/PhysRevD.78.032006
\bibitem{Q&A2004}
J.-S. Wu, W.-T. Ni, S.-J. Chen (Q \& A Collaboration), Class. Quantum Grav. {\bf21} (2004) S1259.
https://doi.org/ \!\!10.1088/0264-9381/21/5/130
\bibitem{Q&A2010}
H.-H. Mei {\em et al.} (Q \& A Collaboration), Modern Phys. Lett. A {\bf25} (2010) 98.
https://doi.org/ \!\!10.1142/S0217732310000149
\bibitem{BMV2004}
C. Rizzo, C. Robilliard, G. Tr\'enec (BMV Collaboration), J. Phys. IV France {\bf119} (2004) 97.
https://doi.org/ \!\!10.1051/jp4:2004119013
\bibitem{BMV2014}
A. Cad\`ene {\em et al.} (BMV Collaboration), 
Eur. Phys. J. D {\bf68} (2014) 16.
https://doi.org/ \!\!10.1140/epjd/e2013-40725-9
\bibitem{OSQAR2006}
P. Pugnat {\em et al.} (OSQAR collaboration), ``Optical Search for QED vacuum magnetic birefringence, Axions and photon Regeneration (OSQAR)'', Report No. CERN-SPSC-2006-035, SPSC-P-331.
https://cdsweb.cern.ch/record/997763/files/spsc-2006-035.pdf
\bibitem{PVLAS2013}
F. Della Valle {\em et al.} (PVLAS Collaboration), 
New J. Phys. {\bf15} (2013) 053026.
https://doi.org/ \!\!10.1088/1367-2630/15/5/053026
\bibitem{OVAL2017}
X. Fan {\em et al.} (OVAL Collaboration), 
Eur. Phys. J. D {\bf71} (2017) 308.
https://doi.org/ \!\!10.1140/epjd/e2017-80290-7
\bibitem{IntrinsicNoise}
G. Zavattini {\em et al.} (PVLAS Collaboration), 
Eur. Phys. J. C {\bf78} (2018) 585.
https://doi.org/ \!\!10.1140/epjc/s10052-018-6063-y
\bibitem{Rossi2003}
L. Rossi, IEEE Trans. Appl. Supercond. {\bf13} (2003) 1221.
https://doi.org/10.1109/TASC.2003.812639
\bibitem{RotatingWavePlates}
G. Zavattini {\em et al.}, Eur. Phys. J. C {\bf76} (2016) 294.
https://doi.org/ \!\!10.1140/epjc/s10052-016-4139-0; {\em ibid.} {\bf77} (2017) 873.
https://doi.org/ \!\!10.1140/epjc/s10052-017-5448-7
\bibitem{CERN_LOI}
R. Ballou {\em et al.} (VMB@CERN Collaboration), ``Letter of Intent to Measure Vacuum Magnetic Birefringence: The VMB@CERN Experiment'', Tech. Rep. CERN-SPSC-2018-036/SPSC-I-249, CERN, Geneva, 2018.
https://cds.cern.ch/ \!\!record/2649744
\bibitem{PBC2020}
J. Jaeckel, M. Lamont, C. Vall\'ee, 
Nat. Phys. {\bf16} (2020) 393. https://doi.org/ \!\!10.1038/s41567-020-0838-4
\bibitem{Bouchiat1982}
M.A. Bouchiat, L. Pottier, 
Appl. Phys. B {\bf29} (1982) 43.
https://doi.org/ \!\!10.1007/BF00694368
\bibitem{Carusotto1989}
S. Carusotto {\em et al.}, 
Appl. Phys. B {\bf48} (1989) 231.
https://doi.org/ \!\!10.1007/BF00694350
\bibitem{Micossi1993}
P. Micossi {\em et al.} (PVLAS Collaboration), 
Appl. Phys. B {\bf57} (1993) 95.
https://doi.org/ \!\!10.1007/BF00425990
\bibitem{Brandi1997}
F. Brandi {\em et al.} (PVLAS Collaboration), 
Appl. Phys. B {\bf65} (1997) 351.
https://doi.org/ \!\!10.1007/s003400050283
\bibitem{Zavattini2006}
G. Zavattini {\em et al.} (PVLAS Collaboration), 
Appl. Phys. B {\bf83} (2006) 571.
https://doi.org/ \!\!10.1007/s00340-006-2189-y
\bibitem{FiveNine}
FiveNine Optics, Boulder, CO (USA), private communication.
\bibitem{TemperatureDependence}
S.M. Etzel, A.H. Rose, C.M.Wang, Appl. Opt. {\bf39} (2000) 5796.
https://doi.org/10.1364/AO.39.005796
\bibitem{Ghosh1999}
G. Ghosh, Opt. Comm. {\bf163} (1999) 95.
https://doi.org/ \!\!10.1016/S0030-4018(99)00091-7
\bibitem{Chen2007}
S.-J. Chen, H.-H. Mei, W.-T. Ni (Q \& A Collaboration), 
Modern Phys. Lett. A {\bf22} (2007) 2815.
https://doi.org/10.1142/S0217732307025844
\bibitem{Mei2009}
H.-H. Mei {\em et al.} (Q \& A Collaboration), 
Chem. Phys. Lett. {\bf471} (2009) 216.
https://doi.org/10.1016/j.cplett.2009.02.048
\bibitem{CPL2014}
F. Della Valle {\em et al.} (PVLAS Collaboration), 
Chem. Phys. Lett. {\bf592} (2014) 288.
https://doi.org/10.1016/j.cplett.2013.12.049
\bibitem{Pugnat_private}
P. Pugnat, private communication.
\end{thebibliography}
\end{document}